\documentclass[fleqn,usenatbib]{mnras}
\usepackage{graphicx}
\usepackage{natbib}
\usepackage{amssymb}
\usepackage{amsmath}
\usepackage{array,booktabs}
\usepackage{mathptmx}
\usepackage{booktabs}
\usepackage{hyperref}
\usepackage{float}
\usepackage{subfig}

\newcommand{\msun}{\,{\rm M_{\odot}}}
\newcommand{\cm}{\,{\rm cm}}
\newcommand{\s}{\,{\rm s}}	
\newcommand{\erg}{\,{\rm erg}}
\newcommand{\teng}{\,{t_{\rm eng}}}
\newcommand{\tengmax}{\,{t_{\rm eng,max}}}

\newcommand{\MBH}{\,{M_{\rm BH}}}
\newcommand{\vrotz}{\,{\omega_0}}
\newcommand{\vrot}{\,{\omega_{\rm rot}}}
\newcommand{\vrotzmin}{\,{\omega_{0,min}}}
\newcommand{\vrotmin}{\,{{\omega}_{\rm rot,min}}}
\newcommand{\rcirc}{\,{r_{\rm circ}}}
\newcommand{\BH}{\,{\rm BH}}

\newcommand{\tbobs}{\,{t_{b,\rm obs}}}

\newcommand{\Bhmin}{\,{B_{h,\rm min}}}
\newcommand{\Phmin}{\,{\Phi_{h,\rm min}}}
\newcommand{\Lmin}{\,{L_{\rm min}}}
\newcommand{\Lminz}{\,{L_{\rm min,0}}}
\newcommand{\Lmax}{\,{L_{\rm max}}}

\newcommand{\dd}[1]{_{_{\rm #1}}}
\newcommand{\risco}{\,{r\dd{\rm ISCO}}}

\usepackage{xcolor}
\usepackage[normalem]{ulem}

\title[Collapsar simulations reveal a range of transients]{Black hole to breakout: 3D GRMHD simulations of collapsar jets reveal a wide range of transients}

\author[O. Gottlieb et al.]{
	Ore Gottlieb$^{1}$\thanks{ore@northwestern.edu},
	Aretaios Lalakos$^{1}$,
	Omer Bromberg$^{2}$,
	Matthew Liska$^{3}$,
	Alexander Tchekhovskoy$^{1}$
	\\
	$^{1}${Center for Interdisciplinary Exploration \& Research in Astrophysics (CIERA), Physics \& Astronomy, Northwestern University, Evanston, IL 60201, USA}\\
	$^{2}${School of Physics and Astronomy, Tel Aviv University, Tel Aviv 69978, Israel}\\
	$^{3}${Institute for Theory and Computation, Harvard University, 60 Garden Street, Cambridge, MA 02138, USA; John Harvard Distinguished Science and ITC}\\
Fellow
}
\pubyear{2021}
\begin{document}
\label{firstpage}
\pagerange{\pageref{firstpage}--\pageref{lastpage}}
\maketitle	
\begin{abstract}
We present a suite of the first 3D GRMHD collapsar simulations, which extend from the self-consistent jet launching by an accreting Kerr black hole (BH) to the breakout from the star. We identify three types of outflows, depending on the angular momentum, $ l $, of the collapsing material and the magnetic field, $ B $, on the BH horizon: (i) {\it subrelativistic outflow} (low $ l $ and high $ B $), (ii) {\it stationary accretion shock instability} (SASI; high $ l $ and low $ B $), (iii) {\it relativistic jets} (high $ l $ and high $ B $). In the absence of jets, free-fall of the stellar envelope provides a good estimate for the BH accretion rate. Jets can substantially suppress the accretion rate, and their duration can be limited by the magnetization profile in the star. We find that progenitors with large (steep) inner density power-law indices ($ \gtrsim 2 $), face extreme challenges as gamma-ray burst (GRB) progenitors due to excessive luminosity, global time evolution in the lightcurve throughout the burst and short breakout times, inconsistent with observations. Our results suggest that the wide variety of observed explosion appearances (supernova/supernova+GRB/low-luminosity GRBs) and the characteristics of the emitting relativistic outflows (luminosity and duration) can be naturally explained by the differences in the progenitor structure. Our simulations reveal several important jet features: (i) strong magnetic dissipation inside the star, resulting in weakly magnetized jets by breakout that may have significant photospheric emission and (ii) spontaneous emergence of tilted accretion disk-jet flows, even in the absence of any tilt in the progenitor.
\end{abstract}


\section{Introduction}

The emergence of long duration gamma-ray bursts (GRBs) in active regions of star-forming galaxies \citep{Bloom2002}, links them to the death of massive stars \citep[e.g.,][]{Galama1998,Modjaz2006}, thus associating GRBs with core-collapse supernovae (CCSNe).
The collapsar model \citep{MacFadyen1999} suggests that during the collapse of a massive star, the iron core collapses to form a Kerr black hole (BH). Subsequently, owing to fast rotation an accretion disk forms, allowing the extraction of energy from the BH in the form of Poynting flux, which powers a relativistic jet that punches through the star and generates the GRB signal after it breaks out of the star.  
Wolf-Rayet (WR) stars have been marked as promising progenitors of GRBs thanks to their strong winds, which can deplete their hydrogen shells, thereby mitigating the journey of the jet through the stellar envelope \citep{Woosley1993}.
Indeed, to date, all SNe observed in coincidence with GRBs have been suggested to be of type Ic, associated with WR stars \citep[e.g.,][]{Cano2017}, and implying that depletion of the helium envelope may be required to allow GRBs to emerge.

As the progenitor star sheds its outer envelope, it transports some of its angular momentum to the interstellar medium (ISM) via strong winds. Since it is anticipated that different layers in the star are coupled to each other, at least to some extent, it is expected that the angular momentum loss should result in a slowdown of the stellar core rotation. Furthermore, strong differential rotation generates magnetic torques, which strengthen the coupling and further decelerate the rotation of the Fe core \citep{Heger2005}. The anticipated relatively low angular momentum of WR cores calls into question their ability to form viable accretion disks (\citealt{MacFadyen1999}, but see possible resolutions in e.g., \citealt{Levan2016,Bavera2021}).
The difficulty in forming the accretion disk required for jet launching may also be a key issue in understanding why only some CCSN of type Ic are associated with GRBs while others are not. Another possibility is that jets are launched in some \citep{Modjaz2016} or even in all CCSN Ic progenitor stars, but many fail to break out and produce the observed GRB emission \citep[e.g.,][]{Mazzali2008}, e.g. due to the presence of an extended envelope \citep[e.g.,][]{Margutti2014,Nakar2015}. Indeed, previous works have suggested that a non-negligible fraction of all long GRB jets are choked inside the stellar envelopes, from both observational considerations \citep{Bromberg2011,Sobacchi2017} and numerical perspective \citep{Lazzati2012}.

GRB jets are most likely powered electromagnetically by the rotational energy of a BH through e.g. the Blandford-Znajek (BZ) mechanism \citep[][]{Blandford1977,Narayan1992,Kawanaka2013}.
\citet{Komissarov2009} suggested that within this framework a necessary condition for jet launching, in addition to sufficiently high angular momentum, is that the magnetic energy density has to exceed the plasma energy density in the vicinity of the BH. They verified this argument using 2D relativistic magnetohydrodynamic (RMHD) simulations.
\citet{Tchekhovskoy2015} proposed that once the jet is
 launched, its power initially depends on the available magnetic flux but is eventually limited by the accretion rate onto the BH, which by itself is governed by the free-fall time of stellar material on the BH. Once the accretion rate drops and the jet can no longer support a high Poynting power, the jet engine ceases to operate.
However, this simple analytic model ignores the complex processes that take place in the disk and can affect the magnetic flux on the BH, as well as the jet-disk feedback and jet-star interplay.
To account for these processes General relativistic magnetohydrodynamic (GRMHD) simulations are required. Such simulations enable us to obtain a self-consistent model of the jet launching, with which one can address fundamental questions such as what are the conditions that allow the jet to be launched and whether the engine operates for a sufficiently long time to allow the jet to break out from the progenitor star and produce the GRB.

Another important question concerns the magnetohydrodynamic evolution of the jet material after it is launched and in particular its magnetization upon exiting the star. Understanding this is important for the identification of the energy dissipation mechanism at large distances responsible for the prompt GRB emission.
\citet{Bromberg2016} performed 3D relativistic magnetohydrodynamic (RMHD) simulations of Poynting flux dominated jets formed by the rotation of a magnetized compact object. They found that the jet is subject to magnetic kink instability, which slows down the jet head while it propagates through the stellar envelope. The instability is responsible for substantial magnetic energy dissipation, which takes place at the location where the jet runs into the stellar envelope and pinches, forming a collimation nozzle: this reduces the jet magnetization to $ \sigma \equiv B^2/(4\pi\rho c^2) \sim 1 $, where $ B $ is the comoving magnetic field and $ \rho $ is the comoving mass density. Their simulations however, did not include the effect of gravity on the stellar envelope, considered neither GR effects on the jet launching nor the disk-jet connection.

In this paper we perform the first 3D GRMHD simulations that follow the relativistic jets launched self-consistently by the spin of BHs at the center of collapsing WR-like stars, propagate $\sim 5$ orders of magnitude to the stellar surface and break out. In \S\ref{sec:model}, we present analytic conditions for jet launching and estimate their duration. In \S\ref{sec:setup}, we describe our numerical setup. Motivated by the simulation results, we develop self-consistent estimates of the jet launching conditions, luminosity and duration.
We show that the analytic criteria for jet launching near the BH event horizon are consistent with the simulations (\S\ref{sec:launching}), infer the accretion rate onto the BH (\S\ref{sec:accretion}),
and the consequent jet power and work-time (\S\ref{sec:properties}). In \S\ref{sec:magnetohydrodynamics}, we discuss the jet and disk evolution including the magnetic dissipation in the jet inside the star and the tilt of the disk and jet axis by infalling material. In \S\ref{sec:breakout_emission} we associate the breakout times with types of GRB progenitors and comment on potential implications on the prompt emission. We summarize and discuss the consequences of our results in \S\ref{sec:summary}.

\section{Analytic model overview}\label{sec:model}
\subsection{Jet launching conditions}\label{sec:launching_condition}

A successful launch of a relativistic BZ jet requires an accretion disk to form around the central BH \citep[e.g.,][]{MacFadyen1999}. Thus, an important condition necessitated by the progenitor is sufficiently high angular momentum, so that the infalling gas hits a centrifugal barrier at a circulation radius $\rcirc>\risco$, where $\risco$ is the innermost stable circular orbit around the BH. Instability processes in the disk such as magneto-rotational instability (MRI) can then amplify the magnetic field, increase the coupling between the disk layers and sustain angular momentum transport that is crucial for accretion. Dynamo processes can then transform some of the predominantly toroidal (pointing in the $\phi$-direction) field formed by, e.g., the disk shear or the MRI, into the poloidal field \citep[pointing in the $R$- and $z$-directions; e.g.,][]{Mosta2015,Liska2020}, which facilitates the extraction of rotational energy from the BH as electromagnetic Poynting flux via the BZ process.

For the launching of a two-sided jet, the BZ power depends on the BH spin (assumed to be aligned with the disk) and on the magnetic flux threading the BH horizon at $ r_h $ \citep[e.g.,][]{Tchekhovskoy2011}\footnote{Note that here the normalization is different since we are using both sides of the jet.}:
    \begin{equation}\label{eq:PBZ}
        L \approx \frac{10^{-3}}{c} \Phi_h^2\Omega_h^2f(\Omega_h)\approx 10^{51} M_{\BH,5}^2 B_{h,15}^2a_{-0.1}^2 \erg~\s^{-1}~,
      \end{equation}
where $ Q_x $ denotes the value of the quantity $ Q $ in units of $ 10^x $ times its c.g.s. units, except for $ M_x $ which is given in units of $ \msun $. $ \Phi_h = 4\pi r_h^2|B_h| $ is the integrated magnetic flux on one hemisphere of the BH horizon, $|B_h|$ is the value of the radial contravariant magnetic field on the horizon, $ \MBH $ and $ a $ are the BH mass and spin, $ \Omega_h = ac/(2r_h) $ is the angular velocity at the BH horizon, and $ f(\Omega_h) \approx 1+1.38x^2-9.2x^4 $, where $ x \equiv 0.5a\left(1+\sqrt{1-a^2}\right)^{-1} $.
A second necessary condition for a successful launching of a BZ jet is that the BZ jet power will be sufficient to overcome the accretion power of the infalling material along the jet path \citep{Burrows2007,Komissarov2009}. Prior to the jet launching, matter is free-falling on the BH quasi-spherically with a power on one hemisphere of
    \begin{equation}\label{eq:Eacc}
        \dot{M}c^2 = 4\pi r_g^2 \rho_h \beta^r_h c^3\approx 2\times 10^{51} M_{\BH,5}^2\beta^r_h\rho_{h,7} \erg~\s^{-1}~,
    \end{equation}
where $ \rho_h $ is the mass density on the horizon at the time of the jet launching, and $ \beta^r_h $ is the dimensionless radial velocity of the infalling material at that time, which is expected to be close to unity. Comparing the jet luminosity from Eq. \ref{eq:PBZ} with the accretion power from Eq. \ref{eq:Eacc}, we get a necessary condition for the strength of magnetic flux on the horizon that allows a successful jet launching
\begin{equation}\label{eq:minB}
    \Phi_h \gtrsim \Phmin \approx 7\times10^{27}\frac{\sqrt{\beta^r_h\rho_{h,7}}}{a_{-0.1}}~{\rm G~{\rm cm^2}}~,
\end{equation}
corresponds to minimal magnetic field on the horizon $ \Bhmin \approx 1.4\times 10^{15} $ G for the same normalization of parameters.
Note that since $ a $ and $ \beta^r_h $ are both of order unity, the minimal magnetic field is primarily dictated by the central density of the star, which in turn depends on the stellar mass, radius and density profile.

\subsection{Jet engine work-time}\label{sec:engine_condition}

We assume that the accretion timescale of the stellar envelope is set by the free-fall time, such that a shell consumed by the BH at a time $ t $ is coming from an initial radius
\begin{equation}\label{eq:rff}
    r_0(t) = (2GM_{\BH}t^2)^{1/3}~,
\end{equation}
where we neglect gravitational effects of the accreted shells on each other, and the effect of the jet-cocoon outflows blocking some of the infalling gas.
In the absence of a jet, the gas free-falls onto the accretion disk, whose size might be changing slowly with time.
If the shell originates from an initial density power-law profile $ \rho(r) = \rho_0 r^{-\alpha} $, we can use Eq. \ref{eq:rff} to estimate the mass accumulation rate on the accretion disk, 
\begin{equation}\label{eq:Md}
\dot{M}_d =
\frac{8\pi\rho_0}{3}(2GM_{\BH})^{1-\frac{\alpha}{3}}t^{1-\frac{2\alpha}{3}}~,
\end{equation}
where we approximate the disk size to be constant in time.
To model the accretion rate onto the BH we assume that a fraction of the mass that falls on the disk is lost to winds during the accretion process, obtaining
\citep[e.g.][]{Blandford1999,Tchekhovskoy2015}
\begin{equation}\label{eq:dotM}
    \dot{M} = \left(\frac{r_w}{r_d}\right)^{\beta_l}\dot{M}_d~,
\end{equation}
where $ r_w$ is the characteristic radius from which the wind is ejected,   $r_d$ is the outer radius of the accretion disk and $\beta_l$ is a constant that controls the strength of the mass loss rate.

The jet can operate as long as the accretion rate onto the BH is high enough and efficiently converted to Poynting flux to sustain the necessary power for a jet launching. High efficiency is naturally obtained when the disk is in a magnetically arrested (MAD) state \citep[e.g.][]{Tchekhovskoy2011}, where the magnetic pressure that supports the jet is of the order of the infalling mass ram pressure. We therefore first assume a MAD state with a accretion to jet energy conversion efficiency of
\begin{equation}\label{eq:eta}
\eta = \frac{1}{\dot{M}c}\int_{r_h,\sigma_h>1}{(T^{r}_t-\rho c^2)dA_\Omega} = \frac{L}{\dot{M}c^2} \sim 1~,    
\end{equation}
where $T^{r}_t$ is the total energy flux in the radial direction, $dA_\Omega$ is a differential area of a solid angle element and the integration is done on the BH horizon. To distinguish between the jet and winds on the horizon, the integration considers only elements that maintain magnetization on the horizon of $ \sigma_h > 1 $. With this definition $L$ is the available jet power that can preform work.
Note that in our definition of $ \eta $, all quantities are instantaneous and $ \eta $ may temporarily deviate from unity, however when averaged over $ \gtrsim 10 $ ms, it is of order unity.
In this case the accretion rate reflects the jet luminosity and can be used to obtain an upper limit on the work-time of the jet engine, $ \tengmax $. For that purpose, we calculate the time interval between the time 5\% of the star was accreted and the time at which 95\% of the star was accreted.
In practice, it is possible that $ \eta $ may drop well below unity at $ t \ll \tengmax $, thereby shutting off the relativistic jet at a time much earlier than $\tengmax $. The efficiency of the jet launching depends on the strength of the poloidal magnetic field threading the BH horizon and on its length scale. Before reaching the BH, the magnetic field is amplified in the accretion disk by MRI and dynamo processes. MRI produces turbulence in the disk, which can disrupt the shape of the magnetic field, reducing its typical length scale and lead to an early engine shutoff, as we show in \S \ref{sec:engine}.

\section{Setup}\label{sec:setup}

To test and calibrate the analytic assumptions in \S\ref{sec:model} we run a suite of GRMHD simulations of self-consistent jet launching in collapsars, using the 3D GPU-accelerated code \textsc{h-amr} \citep{Liska2019}.
For the simulated collapsars we take a WR-like star with a radius $ R_\star = 4\times 10^{10} $ cm and mass $ M_\star \approx 14\msun $. A Kerr BH is placed at the center of the star with a mass  $ M_{\BH} = 4M_\odot $ (not included in $ M_\star $) and a dimensionless spin $ a = 0.8 $. The simulation begins with the gravitational collapse of the star. We monitor the disk formation and the subsequent jet launching and follow its propagation through the star until it breaks out from the stellar surface.

\begin{figure}
		\centering
		\includegraphics[scale=0.22]{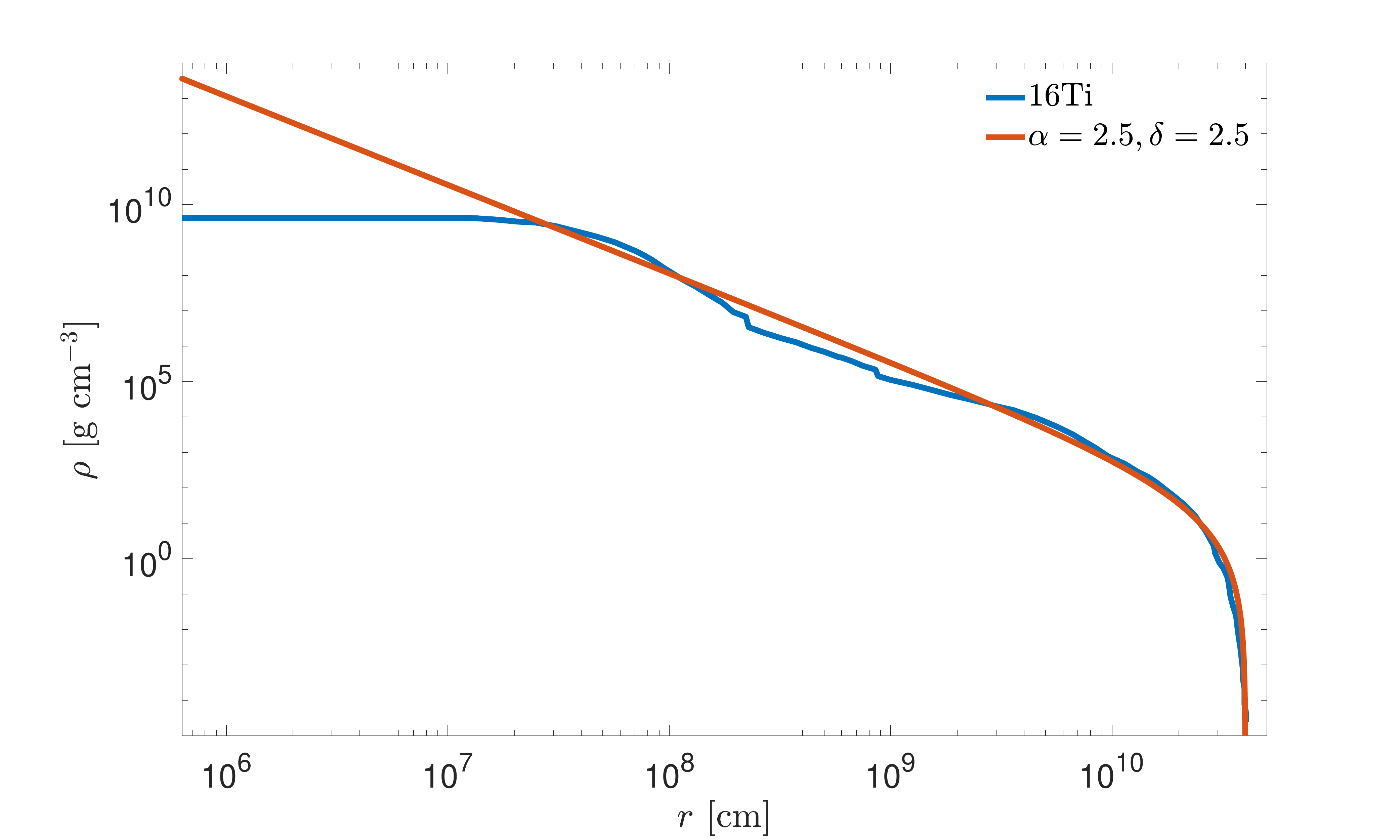}
		\caption[]{
		The simple stellar density profile that we use (Eq. \ref{eq:progenitor}) is consistent with stellar evolution models. Shown here are pre-collapse model 16Ti \cite[][blue]{Woosley2006}, compared with our density profile modeling with $ \alpha = \delta = 2.5 $. Both models feature very similar density profiles over most of the star and hence would result in very similar mass accretion rates (except for factor of a few at very early times).
		}
		\label{fig:16Ti}
\end{figure}

The initial mass profile in the progenitor star just prior to the collapse is modeled as
\begin{equation}\label{eq:progenitor}
    \rho(r) = \rho_0 \left(\frac{r}{r_h}\right)^{-\alpha}\left(1-\frac{r}{R_\star}\right)^\delta~,
\end{equation}
where $ \rho_0 $ is normalized to obtain $\int_{r_h}^{R_\star}\rho(r)d^3r=M_\star$.
This type of progenitor fits stellar evolution models outside of their core. For instance, a commonly used progenitor model 16Ti \citep{Woosley2006} which, like other stellar evolution models, has a pre-collapse uniform density core ($ r \lesssim 2\times 10^7 $ cm) followed by a steep power-law density. Outside the core, the model can be well fitted by Eq. \ref{eq:progenitor} with $ \alpha = \delta = 2.5 $ (see Fig. \ref{fig:16Ti}). We emphasize that the collapse of model 16Ti is indistinguishable from the analytic model at times $ t \gtrsim 10 $ ms, after the collapse of the flat density profile core, whose mass is negligible compared to the mass of the envelope.

We use a spherically symmetric, specific angular momentum profile of the stellar envelope, such that it depends only on the radial coordinate $ \hat{r} $ (except for model $ \alpha1BcLz $ where the rotation is cylindrical, i.e. $ r \rightarrow r{\rm sin}\theta $). The profile is chosen such that it is increasing until $ \sim 70r_g $, and then it becomes constant \citep[see e.g.,][for a similar profile]{Takiwaki2011}:
\begin{equation}
l(r) =
\begin{cases}
\vrotz\left(\frac{r^2}{r_g}\right)^2 & r < 70r_g \\
&\\
\vrotz (70^2r_g)^2 & r > 70r_g
\end{cases} ~,
\end{equation}
where the angular velocity $\vrotz$ is constant.
Last, to set the magnetic field we assume that the star has a distinct magnetic core at the end of its life (i.e. it did not lose its ordered magnetic field by e.g. interlayer mixing), carrying a uniform magnetic field, $ B_0 $, in the $\hat{z}$ direction and adopt an initial dipole-like vector potential outside the core (hereafter core+dipole), which is fully contained within the star, $ r < R_\star $,
\begin{equation}\label{eq:Bprofile}
    A=A_\phi(r,\theta) = \mu\frac{{\rm sin}\theta}{r}\cdot {\rm max}\bigg(\frac{r^2}{r^3 + r_c^3} - \frac{R_\star^2}{R_\star^3 + r_c^3},0\bigg)~,
\end{equation}
where $ \mu \approx B_0r_c^3 $ is the magnetic moment of the uniformly magnetic core, and $ r_c = 10^8 $ cm is the core radius.

\begin{table}
	\setlength{\tabcolsep}{2pt}
	\centering
	\renewcommand{\arraystretch}{2}
	\begin{tabular}{|c|c c c c c|}
    \hline		
	Model & $ \alpha $ & $ {\rm log}\left(\frac{B_h}{\Bhmin}\right) $ & $ {\rm log}\left(\frac{\vrotz}{\vrotzmin}\right) $ & Variation & Outflow
	\\	\hline
    $ \alpha0BcLc $ & 0.0 & 0.0 & 0.0 & canonical & Weak jet
    \\	
    $ \alpha0BshLc $ & 0.0 & 0.0 & 0.0 & $ B(r>r_c) \propto r^{-2} $ & Jet
    \\ 
    $ \alpha0BwLc $ & 0.0 & -0.5 & 0.0 & canonical & SASI
    \\ 
    $ \alpha0BsLc $ & 0.0 & 0.5 & 0.0 & canonical & Jet
    \\
    $ \alpha0BcLs $ & 0.0 & 0.0 & -2.0 & canonical & Subrelativistic
    \\
    $ \alpha0BsLs $ & 0.0 & 1.0 & -2.0 & canonical & Subrelativistic
    \\
    $ \alpha{0.5}BcLc $ & 0.5 & 0.0 & 0.0 & canonical & Jet
    \\	
    $ \alpha{0.5}BsLc $ & 0.5 & 0.5 & 0.0 & canonical & Jet
    \\ 
    $ \alpha1BcLc $ & 1.0 & 0.0 & 0.0 & canonical & Jet
    \\
    $ \alpha1BcLz $ & 1.0 & 0.0 & 0.0 & $ l(r)=l(r{\sin}\theta) $ & Jet
    \\
    $ \alpha1BcLcR_s $ & 1.0 & 0.0 & 0.0 & $ R_\star = 10^9 $ cm & Jet
    \\
    $ \alpha1BtLc $ & 1.0 & 0.0 & 0.0 & $ B(r>500r_g) = 0 $ & Jet
    \\
    $ \alpha1BwLc $ & 1.0 & -1.0 & 0.0 & canonical & SASI
    \\
    $ \alpha1BcL{ms} $ & 1.0 & 0.0 & -1.0 & canonical & Subrelativistic
    \\
    $ \alpha1BcLs $ & 1.0 & 0.0 & -2.0 & canonical & Subrelativistic
    \\ 
    $ \alpha2BcLc $ & 2.0 & 0.0 & 0.0 & canonical & Jet
    \\ 
    $ \alpha2.5BcLc $ & 2.5 & 0.0 & 0.0 & canonical & Jet
    \\ 
    $ \alpha2.5BwLc $ & 2.5 & -1.0 & 0.0 & canonical & SASI
    \\ 
    $ \alpha2.5BcLs $ & 2.5 & 0.0 & -2.0 & canonical & Subrelativistic
    \\ 
    $ \alpha2.5BwL0 $ & 2.5 & -1.0 & $-\infty$ & canonical & None
    \\
    \hline
    \end{tabular}

	\caption{
	Details of the numerical models:
	Model name is composed of the index of $ \alpha $, the strength of the magnetic field on the horizon $ B $ (w: weak, c: canonical, t: truncated, s: strong, and sh: shallow), and angular momentum $ L $ (s: slow, mildly slow: ms, c: canonical, 0: zero, and axisymmetric: z).
	the inner density profile $ \alpha $, logarithmic of the ratio between magnetic flux on the horizon $ B_h $ at the accretion disk formation time and the critical magnetic flux $ \Bhmin $, logarithmic of the ratio between the circularization radius $ \rcirc $ and the gravitational radius $ r_g $, variation of the model with respect to the canonical model, and the outcome of the simulation (see \S\ref{sec:launching}).
		}
\label{tab:models}
\end{table}

We vary $ \alpha, B_0 $ and $ \vrotz $ in our simulations to examine their effect on the jet launching and propagation. We also carry out several simulations where we vary the magnetic field profile and $ R_\star $ to verify our conclusions. The full list of models is given in Tab. \ref{tab:models}.
We restrict the value of $ B_0 $ so that the initial $ \sigma $ everywhere in the star is below unity.
We also keep $ \delta = 3 $ and show analytically that its value does not affect the results considerably.
Similarly, we do not vary the BH spin and mass between simulations as they can only change by a factor of $ \lesssim 2 $ implying a change of up to half an order of magnitude in the jet luminosity (Eq. \ref{eq:PBZ}).
All GRMHD codes need to use density floors to maintain the stability of the numerical scheme. In this spirit, we limit the magnetization to $ \sigma_{\rm max} = 25 $ everywhere in the grid, thus the asymptotic Lorentz factor $ \Gamma_\infty \approx \sigma_{\rm max}$ is much smaller than the values inferred from GRB observations, $ \Gamma_\infty \gtrsim 100 $.

We employ a spherical grid in Kerr-Schild coordinates, use local adaptive time-step, and 2 levels of adaptive mesh refinement (AMR). The refinement criterion is based on the entropy of the fluid in order to properly resolve both the cocoon and the jet.
In simulations where an axisymmetric outflow is generated, we avoid its interaction with the polar axis by directing the rotational axis along the $\hat{x}$ direction, such that the disk forms in the $\hat{y}-\hat{z}$ plane. However, to avoid confusion with the standard convention, in all figures we label the jet axis as the conventional $ \hat{z} $ direction.
The cells are distributed logarithmically in the radial direction, extending out to $ r = 2R_\star $, and uniformly in the polar and azimuthal directions. The grid resolution at the higest AMR level is $ 1152 \times 288 \times 256 $ in $ \hat{r}-\hat{\theta}-\hat{\phi} $ directions, respectively.
We verify that this resolution is sufficient to resolve the fastest growing MRI wavelength by measuring the Q parameter, defined as the ratio between the MRI wavelength and the proper length of a cell in the azimuthal direction. We find that the typical Q parameter is $ 10^2 - 10^3 $, much larger than $ Q \sim 10 $ which is required to properly resolve the MRI \citep[e.g.,][]{Hawley2011}.

\section{Outflow}\label{sec:launching}

\subsection{GRB jet launching}

One requirement for a successful jet launching is linked to the angular momentum distribution within the star. Depending on whether the rotational profile is such that $ \rcirc $ is smaller or larger than $ \risco $, we find two different regimes of jet launching.
For our angular momentum profile the specific angular momentum is maximum at $ r_i \ge 70r_g $. Therefore, to form an accretion disk, the material at this radius needs to possess a high enough angular momentum so that once the gas reaches $ r \sim \risco $ (after a free-fall time of a couple of $ 10 $ ms), an accretion disk forms and the jet is launched. In general, for a radius-dependent angular velocity, $\vrot $, an accretion disk forms if there is a radius $ r_i$ at which the specific angular momentum satisfies $ r_i^2\vrot(r_i)\gtrsim\risco c$. It then follows that the minimal angular velocity at radius $r_i$ with which a jet can be launched is
\begin{equation}\label{eq:vrotmin}
   \vrotmin(r_i) \approx \frac{\risco c}{r_i^2(\vrotmin)}~.
\end{equation}
The formation of a disk is imperative for a stable jet launching as it requires a significant magnetic amplification, which can only take place in accretion disks.
As we show later, our models suggest that if an accretion disk does not form, the accumulation of magnetic field around the BH can only power a weak jet that operates briefly, since the initial poloidal field is not strong enough. Our simulations also show that if an accretion disk does not form early on, the emerging outflow of the weak jet may inhibit the later formation of a relativistic GRB jet. It also implies that a GRB jet should be launched within a few tens of ms after the initial collapse, as was also found by \citet{Mosta2014}, where the precise time depends on the angular momentum, magnetic field and density profiles.
Thus, hereafter for our analytic analysis we assume that an accretion disk forms during the first $ \lesssim 0.05 $ s, and use our simulation parameters $ r_i = 70r_g $ and $ t_i \approx 20 $ ms as the initial radius and the corresponding free-fall time of the matter that forms the disk.

A second requirement is related to the strength of the magnetic field. Consider a shell at a radius $ r_0 $ having width $ \Delta r_0 \lesssim r_0 $ and mass $ m = 4\pi\rho(r_0) r_0^2 \Delta r_0 $ free-falling onto the BH. Its density on the horizon is $ \rho_h \approx \rho(r_0)(r_0/r_h)^2\Delta r_0/\Delta r_h$, where $ \Delta r_h $ is the shell width on the horizon. Free-fall dictates that $ r \propto t^{2/3} $, implying that the shell width on the horizon increases quasi-linearly as $\Delta r_h/\Delta r_0\propto$ $r_0/r_h $. It follows that the shell density on the horizon also increases quasi-linearly as $ \rho_h \approx \rho(r_0)r_0/r_h $.
Since the jet is launched at $ t=t_i $, it encounters a central density $ \rho_h \approx 100\rho(r_i) $. Plugging this into Eq. \ref{eq:minB}, we find that at the time of the jet launching, the minimal magnetic flux on the horizon which allows a successful launching is
\begin{equation}\label{eq:minBh}
    \Phmin \equiv 6\times10^{27}\frac{\sqrt{\beta^r_{h,-0.2}\rho_5(100r_g)}}{a_{-0.1}}~{\rm Mx}~,
\end{equation}
where the calibration $ \beta^r_h \approx 10^{-0.2}\beta^r_{h,-0.2} $ is set by our simulations which show, as expected, that the infalling material is reaching the BH at radial velocities close to the speed of light.
Equivalently, we can use Eqs. \ref{eq:PBZ} and \ref{eq:Eacc} to associate the density with a minimal BZ jet luminosity,
\begin{equation}\label{eq:Lmin0}
    \Lminz = 1.6\times 10^{51} M_{\BH,5}^2\beta^r_{h,-0.2}\rho_5(100r_g) \frac{\erg}{\s}~.
\end{equation}
Since $ M_{\BH,5} $ cannot be much smaller than unity and GRB observations suggest that $ L \lesssim 10^{52}~{\rm erg~s^{-1}} $ \citep[e.g.,][]{Shahmoradi2015}, it follows from Eq. \ref{eq:Lmin0} that GRB progenitors cannot have $ \rho(100r_g) \gtrsim 10^6~{\rm g~cm^{-3}}$.
Due to the magnetic and density profiles, our magnetization profile peaks at $ \sim 100r_g $.

Our estimates also apply in the case where the angular momentum has a cylindrical profile, such as in model $ \alpha1BcLz $.  Although a cylindrical angular momentum creates a low density funnel on the polar axis, it is only important at $ r \gg r_g $ since the density that the jet encounters upon launching is set by the infalling material onto the disk, on timescales shorter than the time it takes to form a funnel through rotation.

\begin{figure}
		\centering
		\includegraphics[scale=0.215]{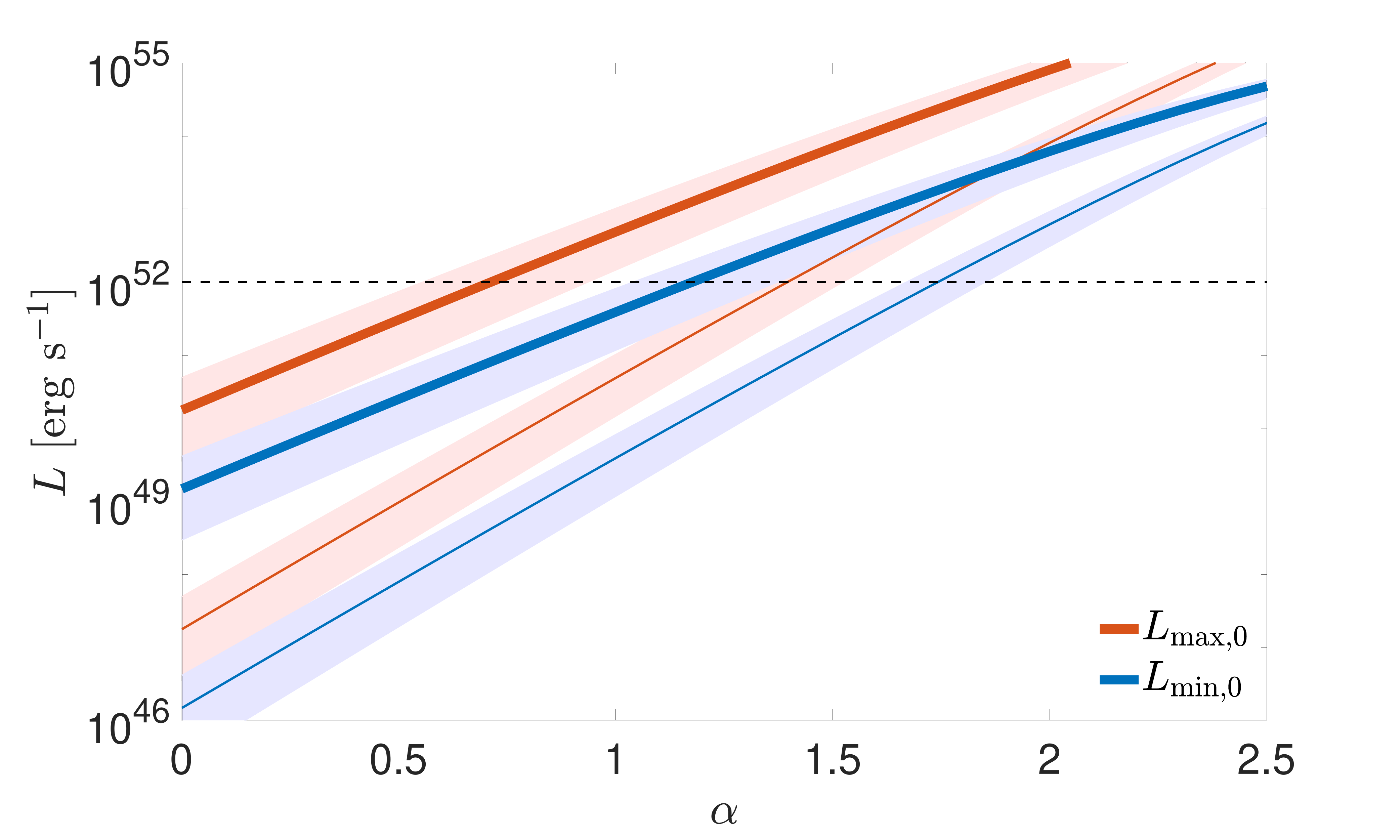}
		\caption[]{Allowed stellar inner density power-law indices $ \alpha $ based on the jet luminosity. Jets in progenitors with $ \alpha \gtrsim 2 $ can only be launched if their luminosity (blue, Eq. \ref{eq:Lmin0}) is $ \gtrsim 10^{52}~{\rm erg~s^{-1}} $ (above black dashed line), in tension with observations of long GRBs \citep[e.g.,][]{Shahmoradi2015}. Also shown is the maximum jet luminosity (red), based on 100\% jet energy production efficiency (jet power $=$ accretion power).
        Luminosities are shown for $ R_\star = 4\times 10^{10} $ cm (thick lines) and $ R_\star = 4\times 10^{11} $ cm (thin lines). The lines are plotted for $ \delta = 3 $, with shaded areas mark the range of $ L $ between $ \delta = 1 $ and $ \delta = 5 $ (see Eq.~\ref{eq:progenitor}).
		}
		\label{fig:L_alpha}
\end{figure}

One can use the maximum central density that allows a jet launching to constrain the inner density profile of GRB progenitors. Assuming for simplicity our canonical $ M_\star = 14\msun $, we find the density at $ 100 r_g $ for each power-law index $ \alpha $, and plot in Fig. \ref{fig:L_alpha} the minimal jet luminosity upon launching, $ L_{\rm min,0} $ (blue) as a function of $ \alpha $. The shaded area represents the variation in $ L_{\rm min,0} $ between values of $ \delta = 1 $ and $ \delta = 5 $. For comparison we also plot in red the maximum luminosity obtained if 100\% of the accretion power at the time of the jet launching, is channeled to the jet ($\eta=1$). The actual power of a successful GRB jet will lie between those two lines. Since the density is linear in  $ M_\star $, which cannot be much smaller than  $ 14 \msun $, a change in the stellar mass does not affect this result to a large extent.
On the other hand, $ R_\star $ may vary by an order of magnitude\footnote{Note that our definition for $ R_\star $ is set by the density profile in Eq. \ref{eq:progenitor}, and does not include the case of a progenitor that is embedded in an extended envelope.} and its effect on the density is superlinear. Thus, we plot the critical luminosity both for $ R_\star = 4\times 10^{10} $ cm (thick lines) and $ R_\star = 4\times 10^{11} $ cm (thin lines).
Fig. \ref{fig:L_alpha} demonstrates that only moderate density profile slopes at the stellar interior ($ \alpha \lesssim 1.5 $) are allowed as they lead to sufficiently low central densities, which support the formation of jets with observed luminosities of $ L \lesssim 10^{52}~{\rm erg~s^{-1}} $.
Jets in progenitors with $ \alpha > 2 $ are ruled out as GRB candidates due to their excessive power. For example, when the density profile index is $ \alpha = 2.5 $, $ \rho(100r_g) \approx 10^9~{\rm g~cm^{-3}} $, and thus such stars can only launch a jet with a luminosity that is much higher than all detected long GRBs.

\begin{figure*}
		\centering
		\includegraphics[scale=0.08]{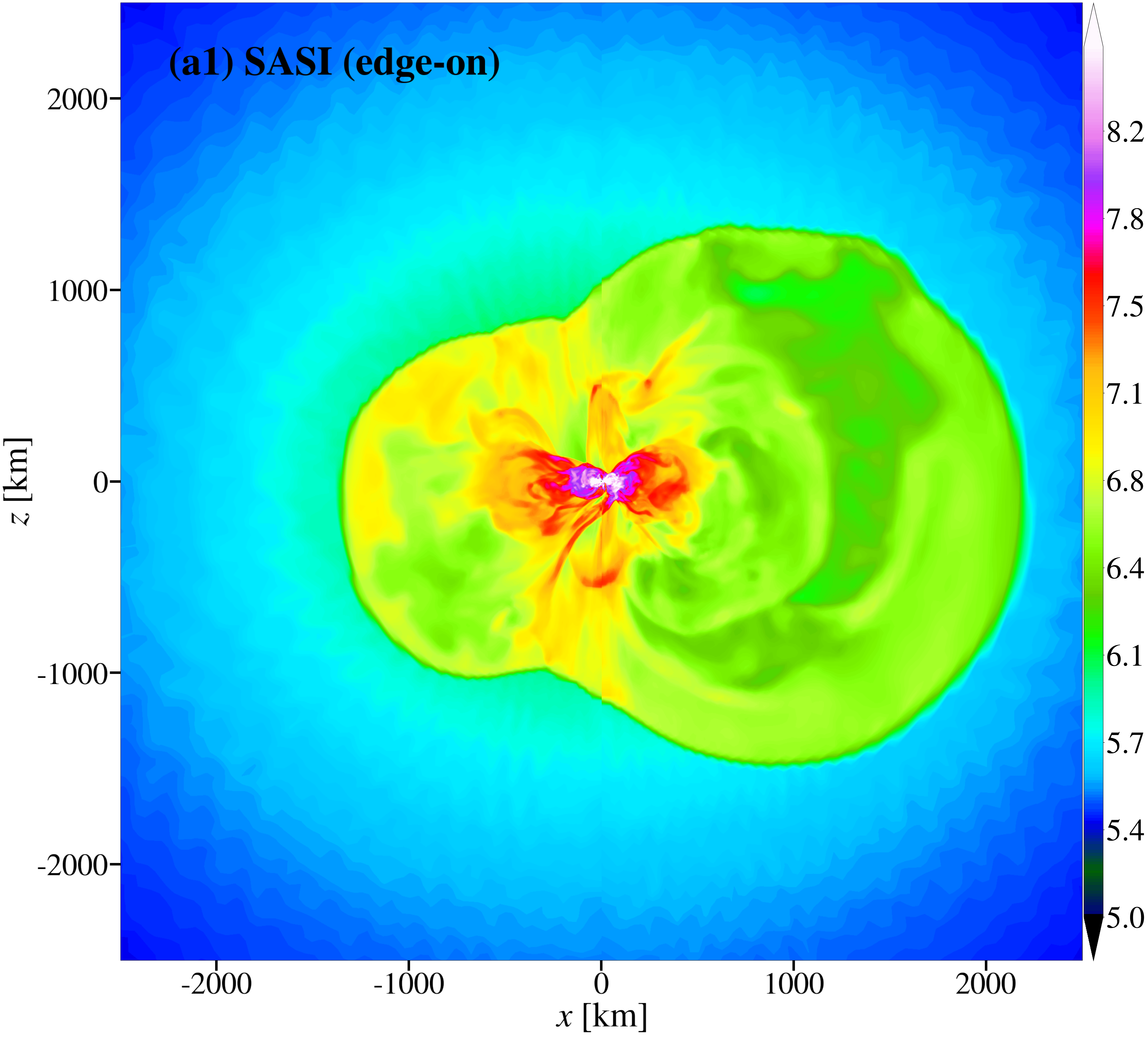}
		\includegraphics[scale=0.08]{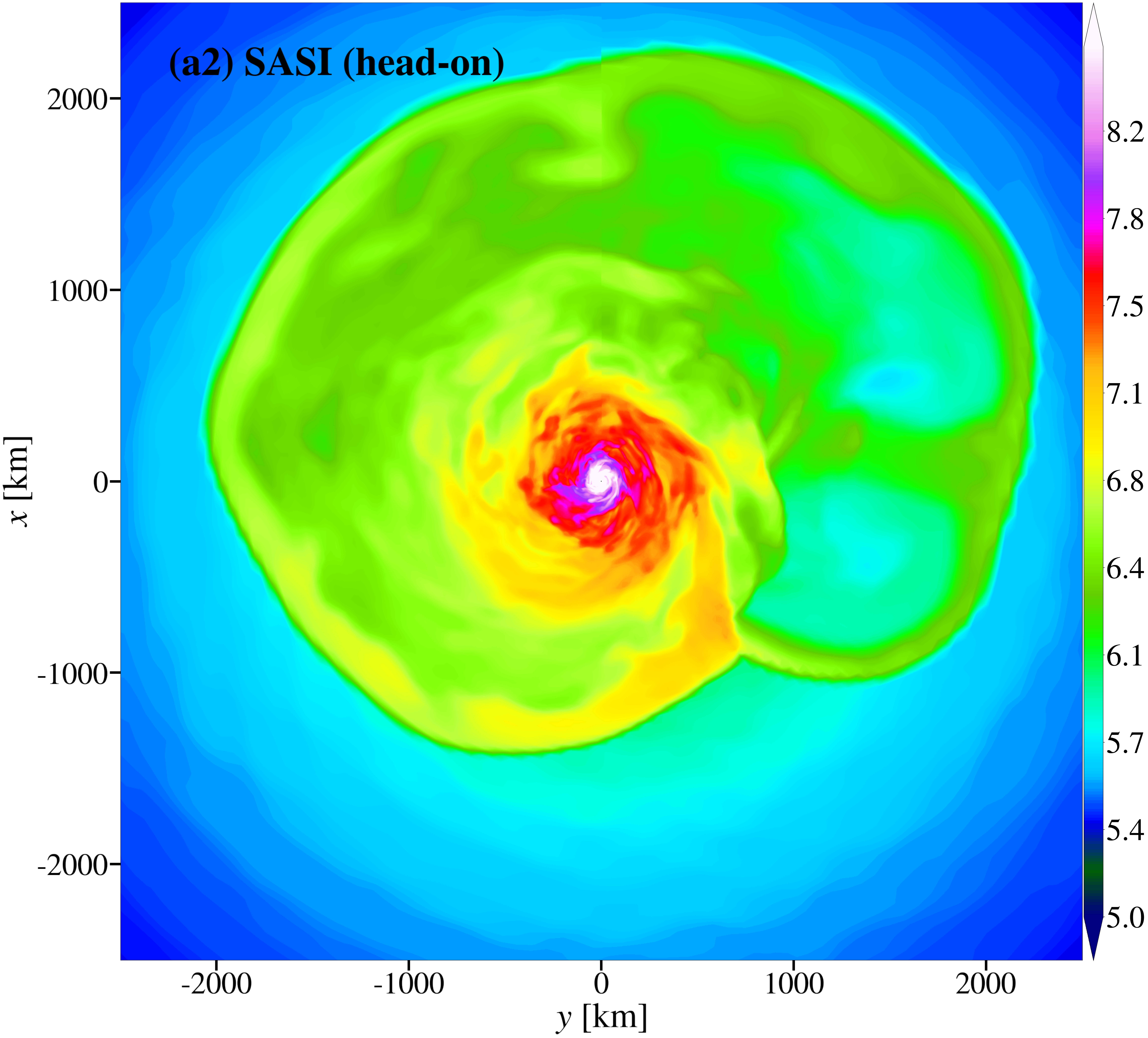}
		\includegraphics[scale=0.098]{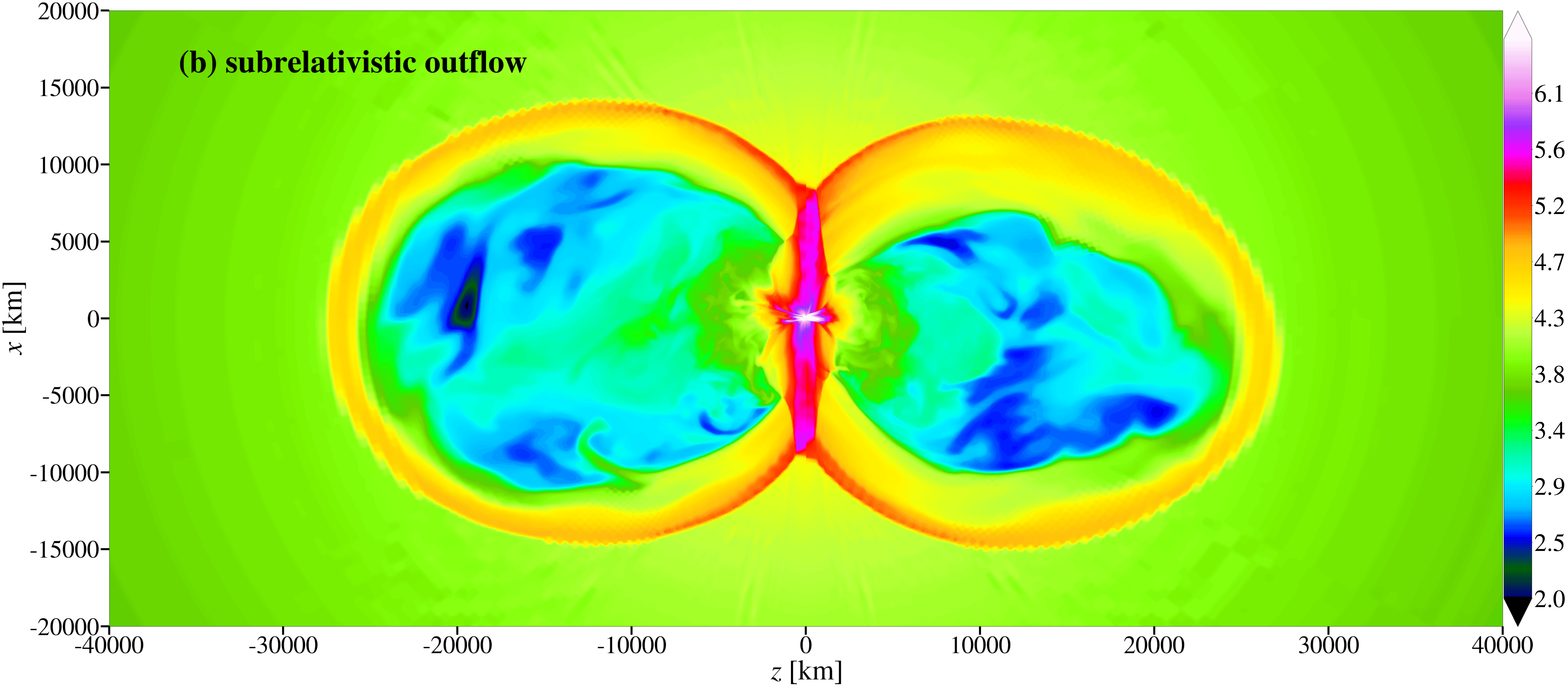}
		\includegraphics[scale=0.109]{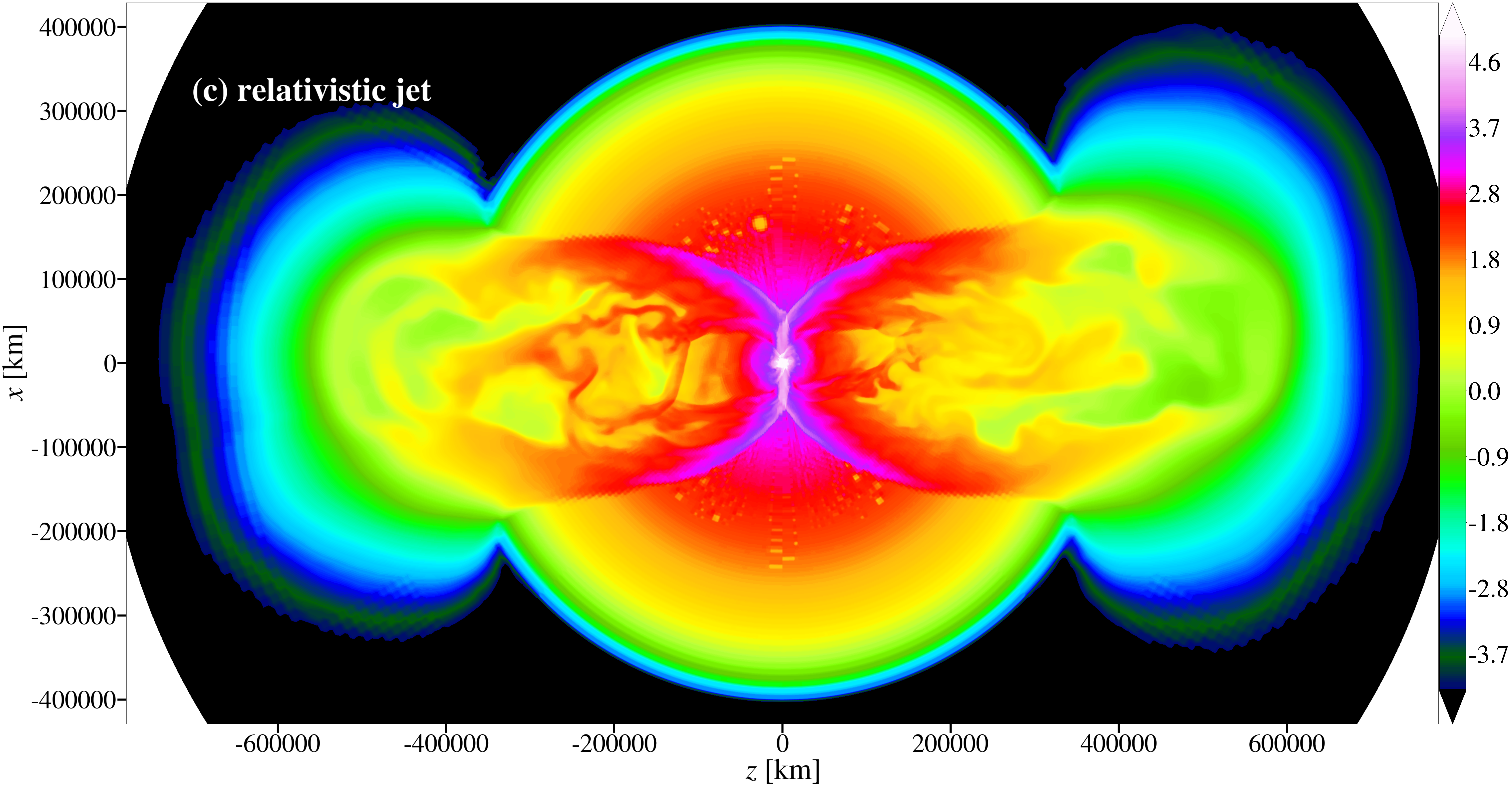}
		\caption[]{A variety of outflows from different progenitors, based on Tab. \ref{tab:overview}.
		Displayed logarithmic mass density maps of: a) SASI: high specific angular momentum to form an accretion disk, and weak magnetic flux which is insufficient to launch a jet. Shown is an edge-on view (a1) and head-on view (a2), taken from model $ \alpha1BwLc $, 0.3 s after the collapse; b) subrelativistic outflow: specific angular momentum is too low to form a disk and magnetic field is strong to extract energy from the BH. Shown is an edge-on view from model $ \alpha1BcLs $, 1.2 s after the collapse; c) jet: high specific angular momentum and strong magnetic field support jet launching.
		Shown is an edge-on view when the forward shock reaches $ \sim 2R_\star $ from model $ \alpha1BcLz $, 12 s after the collapse. The central engine shut itself off before the jet breakout and thus the breakout is of the shocked jet material.
		In panels (b,c) the metric is tilted such that in all panels the accretion disk lies on the $ \hat{x}-\hat{y} $ plane.
		Videos are available at \url{http://www.oregottlieb.com/collapsar.html}.
		}
		\label{fig:outcomes}
\end{figure*}

\subsection{Types of outflows}

Following the discussion above, our simulation results can be divided into four types of outflows from the BH horizon, depending on the relation between $ \vrotz, B_h $ and their critical values $ \vrotzmin, \Phmin $ obtained from Eqs. \ref{eq:vrotmin} and \ref{eq:minBh}, respectively. The outflow types are listed in Tab. \ref{tab:overview} (see also Tab. \ref{tab:models}), and illustrated in Fig. \ref{fig:outcomes} showing snapshots of logarithmic density maps of three cases in which an outflow emerges: 

\begin{table}
	\setlength{\tabcolsep}{13pt}
	\centering
	\renewcommand{\arraystretch}{2}
	\begin{tabular}{c|c c}
		
		& $ \rcirc \gtrsim \risco $ & $ \rcirc < \risco $
		\\	\hline
        $ \Phi_h \gtrsim \Phmin $ & Relativistic jet & Subrelativistic outflow
        \\	
        $ \Phi_h \lesssim \Phmin $ & SASI & Gravitational collapse
        \\ 
	\end{tabular}

	\caption{The four possible outcomes of outflow from the BH horizon, depending on the specific angular momentum of the star which determines if an accretion disk forms, and magnetic field in the star which dictates whether the jet power is sufficient to overcome the infalling accretion power.
		}
\label{tab:overview}
\end{table}

i) {\it SASI} ($ \rcirc \gtrsim \risco $ and $ 
\Phi_h \lesssim \Phmin $):
a rapid rotation with $ \vrotz \gtrsim 10~{\rm s^{-1}} $ results in the formation of an accretion disk around the BH. When the magnetic field on the horizon is weak, the Poynting power is insufficient to overcome the ram pressure of the infalling material and no jet is launched. Wind from the hot accretion disk collides with the falling material forming a quasi spherical shock around the disk. The continual flow of wind energizes the shock and drives hydrodynamic instabilities, which lead to a behavior similar to that seen in accretion shocks around CCSNe, known as stationary accretion shock instability \citep[SASI;][]{Blondin2003,Blondin2006}.
We stress that the formation of the spherical shock wave in CCSNe is different, as it is driven by the stellar core-bounce and stalled at $ \sim 10^7 $ cm due to neutrino losses and the dissociation of heavy nuclei such as iron into nucleons \citep[see e.g.,][]{Burrows1995,Mezzacappa1998}.
In the absence of neutrino scheme and nuclear physics in our simulations, the shock is not halted at any point, and thus only resembles the early evolution of SASI.
Nevertheless, both types of shocks show a similar behaviour and lead to symmetry breaking of the explosion.
We find that in all simulations in which an accretion disk forms, the spiral mode ($ m =1 $) of the accretion shock grows by nearly axisymmetric accretion (Fig. \ref{fig:outcomes}a1,a2), as was found in 3D simulations of CCSNe \citep[e.g.,][]{Blondin2007,Fernandez2010}.
When the magnetic field on the horizon is too small to power a jet, the SASI-like structure grows unperturbed, as in a case of a SN that is not accompanied by a GRB jet.
We point out that the growth of accretion shock increases the central density, shuttering any hope for a delayed jet launching.

ii) {\it Subrelativistic outflow} ($ \rcirc \lesssim \risco $ and $ \Phi_h \gtrsim \Phmin $):
the small circularization radius places the centrifugal barrier at a radius smaller than $ \risco $ and an accretion disk does not form. However, the infalling magnetic flux can initially thread the spinning black hole and extract some BH spin energy to briefly power a weak jet. In the absence of a disk that can amplify the magnetic field, the flux on the BH escapes via reconnection on the equatorial plane and weakens until the jet launching ceases.
Such an outflow might have been found by \citet{Burrows2007} who described it as a pre-cursor jet.
The outflow structure at later times depends on the density profile. A quasi-spherical subrelativistic outflow emerges if the jet decelerates in a rather flat density profile (Fig. \ref{fig:outcomes}b), thus its breakout may resemble a weak SN explosion that is continuously powered by the inefficient energy extraction from the BH by a weak field. If the core density profile is steep such that the outflow accelerates, it may keep its elongated shape in what may be the source of low-luminosity, soft GRBs via e.g. sub-relativistic shock breakout emission, which may yield a quasi-isotropic signal.

iii) {\it Relativistic jet} ($ \rcirc \gtrsim \risco $ and $ \Phi_h \gtrsim \Phmin $):
the first criterion indicates that the stellar envelope collapses until reaching the centrifugal barrier to form an accretion disk, which drives a BZ jet launching. The second criterion implies that the jet power is sufficient to overcome the ram pressure of the central density and a relativistic jet is successfully launched from the horizon. The jet propagates in the stellar envelope and forms a hot weakly-magnetized cocoon which collimates the jet (Fig. \ref{fig:outcomes}c).
Since in our simulations both the jet and the accretion shock are driven by the instabilities in the disk, they are built over the same timescale such that the growth of the cocoon disrupts the otherwise spiral structure of the accretion shock. We discuss the jet propagation in the envelope in \S\ref{sec:magnetohydrodynamics}, and leave a detailed analysis of the jet-SASI interplay \citep[][]{Nagakura2008} for a future work.

\section{Accretion rate}\label{sec:accretion}

We show that the accretion rate onto the BH can be estimated by a spherical free-falling rate of the stellar envelope with a few caveats: i) we neglect the unbound mass that could be ejected if there is an explosion prior to the onset of the simulation. ii) our gravitational potential is fixed and does not change with the accretion of mass onto the BH.
iii) the presence of the cocoon may deviate the accretion from a spherical free-fall.  
Our simulations are carried out for progenitors with $ M_\star = 14 \msun $ and $ \delta = 3 $. We emphasize that the accretion rate temporal profile is solely dictated by the value of $ \alpha $, as $ M_\star $ and $ \delta $ affect only the normalization through the density constant $\rho_0$. We begin with calculating the free-falling mass flow rate on the disk using Eqs. \ref{eq:rff} and \ref{eq:progenitor}
\small
\begin{equation}\label{eq:Macc}
    \dot{M}_d(t,\alpha) = \frac{8\pi\rho_0r_0(t)}{3R_\star^3t}\left(-r(t)^{5-\alpha}+3R_\star r(t)^{4-\alpha}-3R_\star^2r(t)^{3-\alpha}+R_\star^3r(t)^{2-\alpha}\right)~,
\end{equation}
\normalsize
where $ r_0(t) $ is the original radius from which matter is reaching the disk\footnote{At $ t \lesssim 10 $ ms there is no disk to stall the matter from reaching the BH and at $ t \gg 10 $ ms, $ r_0 $ is much larger than the size of the disk, so the free-falling time to the BH and the disk is similar.} at time $ t $.

\begin{figure}
		\centering
		\includegraphics[scale=0.25]{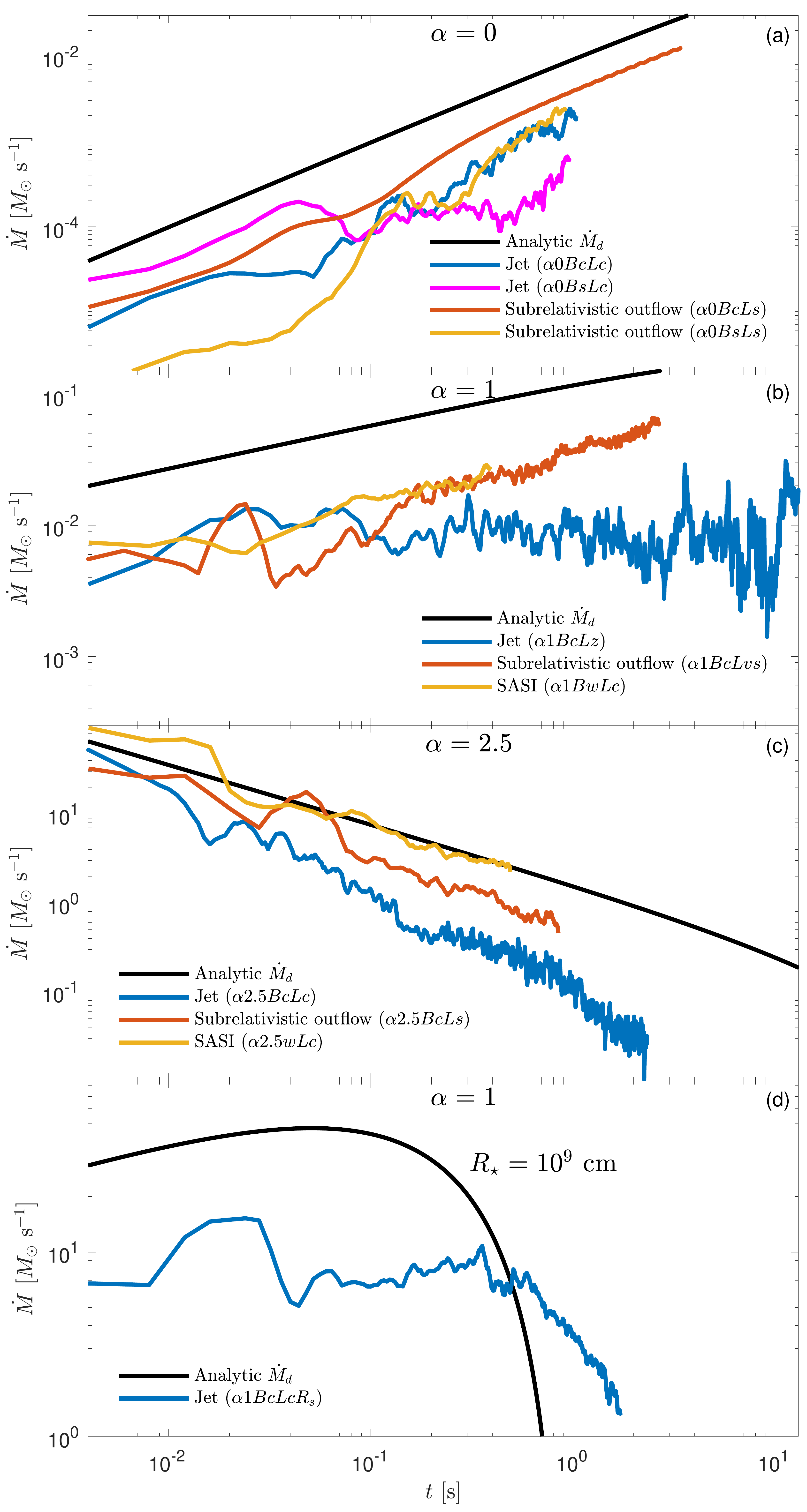}
		\caption[]{
		Analytic model for free-fall mass accretion rate (black) provides an upper limit to $\dot M$ from our simulations (colored lines). The power-laws are in agreement in the absence of the jet, $ \dot{M} \propto t^{(3-2\alpha)/3}$, whereas when a jet is present, $ \dot{M} \propto t^{(2-2\alpha)/3}$.
		(a)-(c): our canonical progenitor is considered with a variety of density profiles, $ \alpha = 0,1,2.5$.
		(d): the accretion rate for a small star, $ R_\star = 10^9 $ cm, showing that the time at which the accretion rate drops is in agreement between the analytic and numerical results.
		}
		\label{fig:accretion_rate}
\end{figure}

Fig. \ref{fig:accretion_rate} shows the accretion rate onto the BH, $ \dot{M} $, as found in the simulations (colored lines), for a variety of $ \alpha $, $ B_0 $ and $ \vrotz $, compared with the analytic expression for the mass flow rate on the disk in Eq. \ref{eq:Macc} (black lines). In the top three panels, which show the results for progenitors with $ R_\star = 4\times 10^{10} $ cm, the analytic accretion rate evolves as a power-law throughout the entire simulation. In simulations where there is no jet (subrelativistic outflow/SASI in red and yellow lines) the measured accretion rate is in a remarkable agreement with the analytic prediction of free-fall time, both in the slope and the normalization which differs by up to a factor of $ \lesssim 3 $.
This factor can be explained by a combination of an aspherical accretion in the case of a subrelativistic outflow whose cocoon blocks some of infalling matter from reaching the disk, and winds which lead to a loss of some of the accreting mass onto the accretion disk before reaching the BH as discussed in \S\ref{sec:engine_condition}.
If the factor is attributed to winds from the disk, an agreement between the analytic accretion rate given in Eq. \ref{eq:dotM} and the simulated ones can be obtained by taking the size of the accretion disk $ r_d \approx 100r_g $ from the simulation, and using $ r_w \approx 10r_g $,  and $ \beta_l \approx 0.5 $, consistent with the values found in other numerical studies \citep[e.g.,][]{McKinney2012}.

When a jet is present (blue lines), the cocoon expands fast (in contrast to the subrelativistic outflow) and suppresses some fraction of the free-falling gas into the disk. The main gas component that ultimately powers the accretion flows along the lower sides of the expanding cocoons of the two jets towards the equator, where it hits a strong accretion shock and is stalled.
Thus, it takes longer to reach the disk, depending on the shape of the cocoon and on the location of the accretion shock at the equator, which evolves as $ t^{\frac{6-\alpha}{5-\alpha}} $ \citep{Gottlieb2021c}.
However, the accretion rate is not only linked to the cocoon width. First, while part of the shocked gas in the cocoon becomes unbound, some of the cocoon gas does fall back onto the disk. Second, interaction between the cocoon and the counter cocoon on the equator drives shocks which render the pressure in the cocoon nonuniform.
Therefore, we do not provide an analytic solution to the influence of the cocoon on the accretion rate, but study the accretion rate in this case from the numerical simulations.
Our simulations show that as long as the accretion originates in the inner density power-law segment, the accretion rate scales as $ \dot{M} \sim t^{(2-2\alpha)/3}$ (slower than the analytic estimate of $ \dot{M} \propto t^{(3-2\alpha)/3}$).

Finally, our simulations with $ R_\star = 4\times 10^{10} $ cm do not last long enough to reach the expected drop in the accretion rate at $ t \gtrsim 100 $ s, which takes place when the outer layers of the envelope are accreted. Thus, in order to examine the change in the accretion rate at late times, we also simulate a smaller progenitor with $ R_\star = 10^9 $ cm, while keeping $ M_\star = 14 \msun $ (model $ \alpha1BcLcR_s$). In this progenitor, as shown in Fig. \ref{fig:accretion_rate}d, the drop in the accretion rate should take place at $ t \approx 0.5 $ s according to Eq. \ref{eq:Macc}, enabling us to explore the final phase of the accretion.
We first note that unlike our canonical sized progenitors, here the sharp decline in the stellar edge affects the density profile already at early times, and thus the initial accretion is roughly, but not fully, consistent with the arguments above for jets.
One can see that the accretion rate has a clear break in its power-law behavior at a time that is consistent with the analytic expectation and it starts falling as $ t^{-2} $ \citep[similar to what was found in the simulations of][]{Christie2019}. We conclude that the accretion duration is also consistent with the analytic estimate.

\section{Jet work-time \& luminosity}\label{sec:properties}

Using the accretion rate behavior, one can infer two fundamental jet quantities: the central engine work-time and the jet luminosity. The jet launching duration, $ \teng $, is governed by the accretion rate, magnetization and by the magnetic field profile of the infalling material.
The maximum luminosity is simply the total accreted power, $\Lmax \equiv \dot{M}c^2 $, as shown in Fig. \ref{fig:L_alpha} (red line) at the time of the jet initial launching, where $ \dot{M} $ is calculated under the assumption of free-fall time\footnote{Since this is an upper limit estimate, we ignore the numerical correction of a factor of a few between $ \dot{M} $ and the free-fall approximation.}. The true luminosity depends on the conversion efficiency of accreted mass flow to jet power, $ \eta $ as $ L = \eta\Lmax $.

\subsection{Engine activity}\label{sec:engine}
\begin{figure}
		\centering
		\includegraphics[scale=0.22]{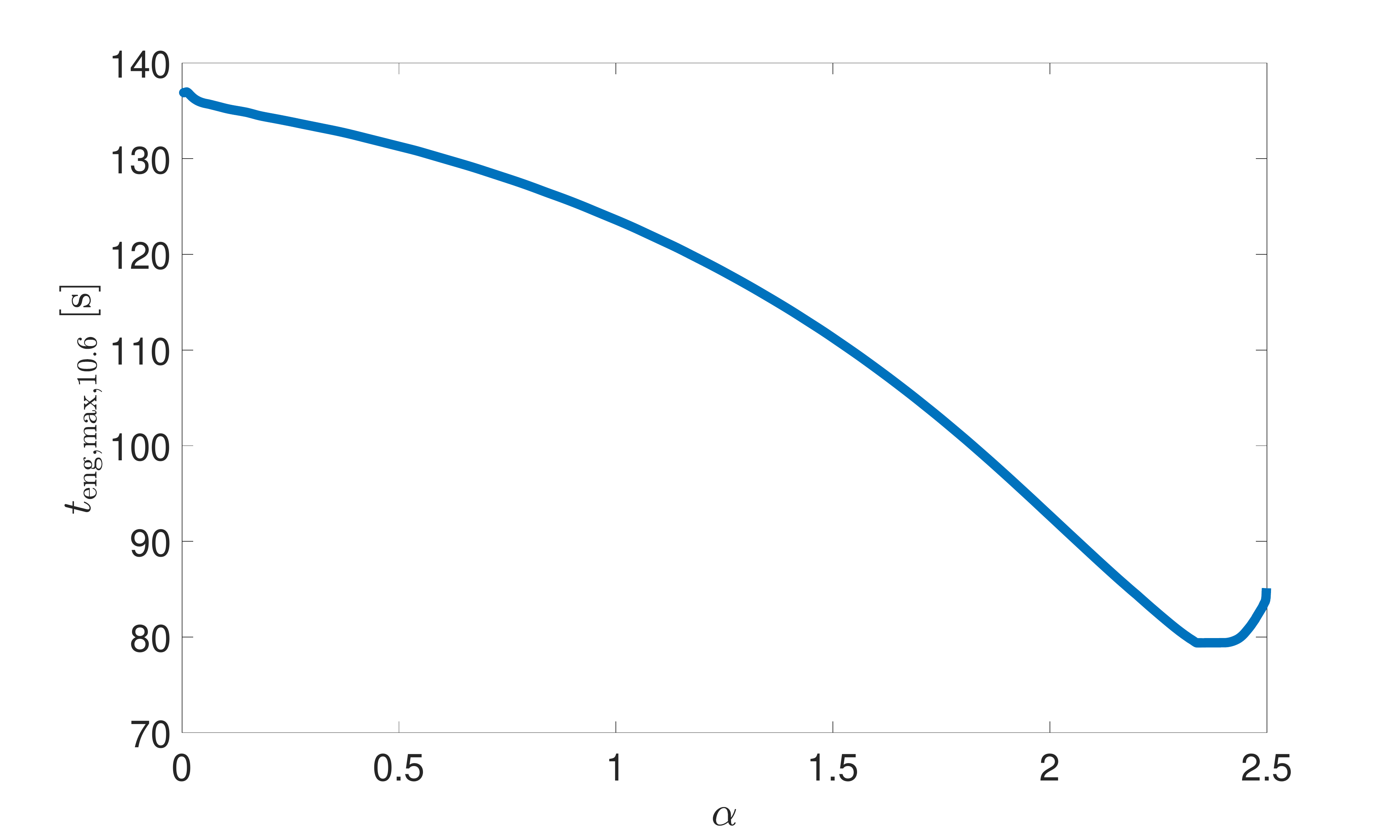}
		\caption[]{An upper limit to the central engine activity time, determined by the free-fall time of the stellar upper envelope, for $ R_\star = 4\times 10^{10} $ cm.
		}
		\label{fig:worktime}
\end{figure}

To evaluate the maximum possible engine activity time, $ \tengmax $, we follow the method presented in \S\ref{sec:engine_condition}, which connects this time with the time where $90\%$ of the stellar material is accreted.
Since our simulations do not last through the entire stellar collapse, we estimate this time analytically by integrating Eq. \ref{eq:Macc} over the entire accretion time. As stated before, $ \tengmax $ is an upper limit for the jet activity time, and thus that the accretion onto the BH can be approximated as $ \dot{M} = \dot{M}_d $ and that the conversion efficiency to jet power is $\eta=1$.
In Fig. \ref{fig:worktime} we show $ \tengmax $ as a function of $\alpha$ for our canonical progenitor of $ R_\star = 4\times 10^{10}~\cm$, denoted as $ t_{\rm eng,max,10.6} $. The obtained values are of the order of $t_{\rm eng,max}\simeq110\pm30$ s for $0\leq\alpha\leq2.5$. Eq. \ref{eq:rff} suggests a scaling relation of engine activity time with stellar radius 
\begin{equation}\label{eq:teng_alpha}
\tengmax(\alpha,R_\star) = \left(\frac{R_\star}{4\times10^{10}~{\rm cm}}\right)^{1.5}t_{\rm eng,max,10.6}(\alpha),
\end{equation}
implying that stars with $ R_\star \approx 4\times 10^{10} $ cm could be more consistent with typical durations of GRB jets, whereas stars with $ R_\star \gtrsim 10^{11} $ cm might be more consistent with ultra-long GRBs.

As we pointed in \S\ref{sec:engine_condition}, efficient conversion of accretion energy to jet power strongly depends on the conditions of the magnetic fields at the vicinity of the BH and may result in $ \teng \ll \tengmax $. In particular the curvature radius of the poloidal field component needs to be larger than a few tens $ r_g $ in order to efficiently convert the BH rotational energy to Pointing flux \citep[e.g.][]{Chashkina2021}. 
The initial field in our simulations consists of an inner core with a uniform vertical field at $ r \lesssim 10^8$ cm surrounded by an outer dipole field. To control the magnetization we modify the density normalization $\rho_0$ and power-law index, $ \alpha $ under the constraint that $ \sigma < 1 $ everywhere inside the star. The resultant magnetization profile holds $\sigma\propto r^{\alpha} $ inside the core and $ \sigma\propto r^{\alpha-6} $ outside of it.
Accretion of such a large scale field builds enough magnetic flux on the BH horizon to trigger a short jet launching episode at the onset of the simulation. 
However, this flux does not survive as it leaks out over dynamical time scales, due to magnetic reconnections that take place on the equatorial plane. The continuous powering of the jet requires a buildup of the flux by magnetic field coming from the accretion disk, where the length scale of this field is set by various instability processes in the disk, such as MRI, which can generate turbulent motions and disrupt the initial field configuration. 

\begin{figure}
		\centering
		\includegraphics[scale=0.28]{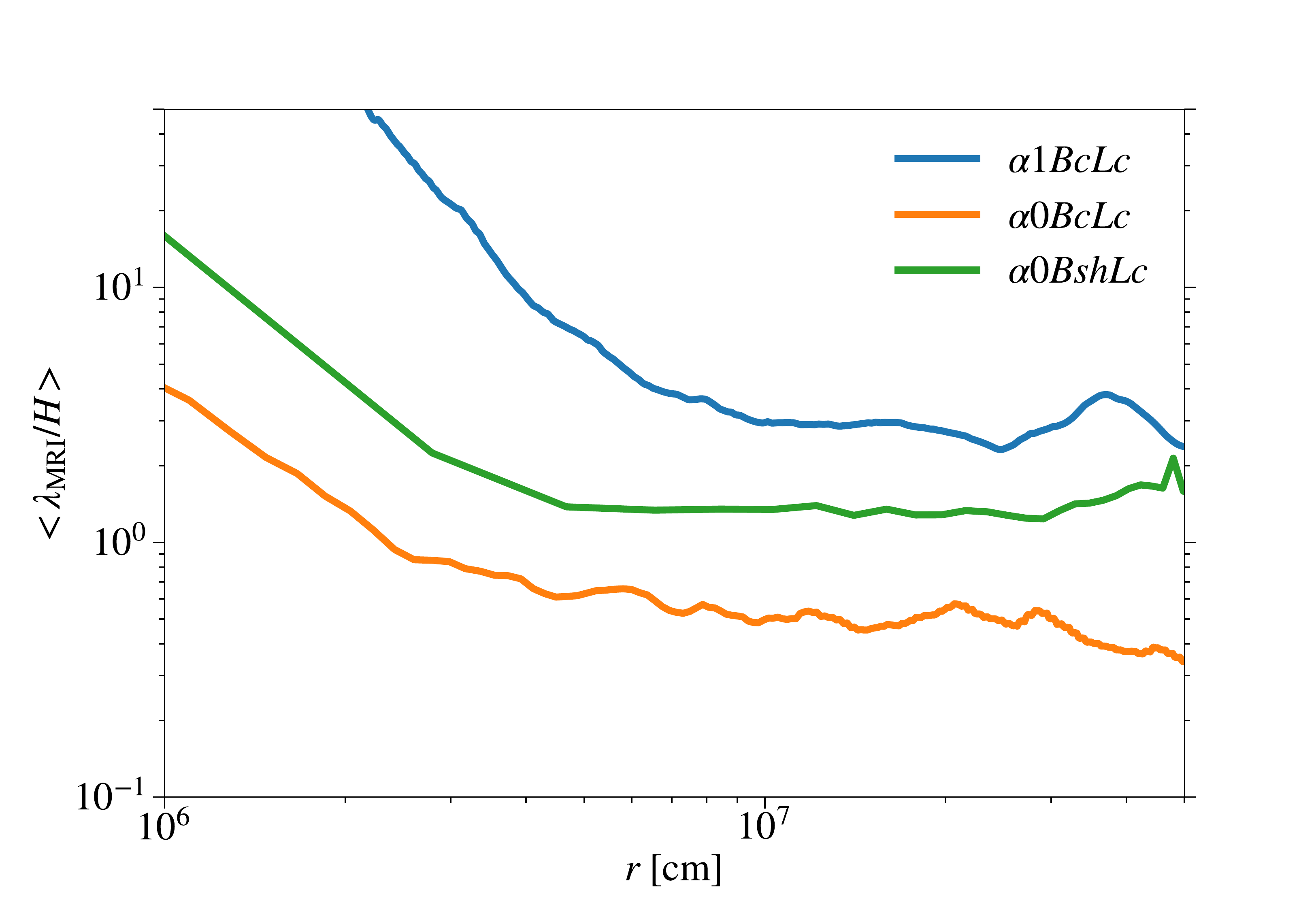}
		\caption[]{Accretion disk magnetization determines its ability to retain the large-scale vertical magnetic flux. Models $ \alpha1BcLc $ (blue) and $ \alpha0BshLc $ (green), which start with a shallower magnetization profile, either due to steeper density profile ($ \alpha1BcLc $) or shallower magnetic field profile ($ \alpha0BshLc $), end up with a higher magnetization of the disk and the vertical MRI wavelength exceeding the disk thickness. This prevents the magnetic field from developing small-scale features and diffusing outwards, resulting in a stable jet for $ \sim 3 $ s before the magnetization in the disk drops. In contrast, when the magnetization profile is steeper, such as in model $ \alpha0BcLc $ (orange), low magnetization matter is accreted early on, leading to $ \lambda_{\rm MRI}<H $ and the growth of small scale turbulence in the disk, which disrupt the large scale profile of the magnetic filed.
		Shown are the radial distributions of $<\tilde\lambda>$, the ratio of the wavelength of the fastest growing MRI mode to the disk scale height, weighted by the local density and averaged over the angular direction. 
		The distribution is shown 0.25 s after the collapse in two models.
		Videos of the evolution in time of the disk in both cases are available at \url{http://www.oregottlieb.com/collapsar.html}.
		}
		\label{fig:MRI}
\end{figure}

In order for turbulence to grow in the disk, the fastest growing MRI mode in the $z$ direction must be of the order of (or smaller than) the disk scale height 
$H=c_s/\vrot$, where $c_s$ is the local sound speed and $\vrot$ is the local angular velocity. In ideal MHD this mode is $\lambda_{{\rm MRI}}=2\pi v_{A,z}/\vrot$ \citep[e.g.][]{Masada2008}, where $v_{A,z}=B_z^2/4\pi\rho$ is the Alfven velocity in the $z$ direction. We can thus define the ratio $\tilde{\lambda}=\lambda_{{\rm MRI}}/H\propto\beta_p^{-1/2}$, and use the condition  $\tilde\lambda\lesssim1$ as an approximate threshold for a successful growth of small scale turbulence in the disk that can disrupt the ordered magnetic field structure and facilitate reconnection that destroy the overall flux that goes into the BH.
Fig. \ref{fig:MRI} displays the radial profile of
$\tilde\lambda$, weighted by the mass density at each point and averaged over all angles, which picks out the disk region out of the infalling material.
In the case of a steep magnetization profile (orange line), $\tilde\lambda$ drops below unity after $ t \approx 0.1$ s which corresponds to a free-fall time of matter that is located initially at a radius twice the size of the magnetized core. The initial magnetization in this case scales like $r^{-6}$ outside the core implying that the magnetization of the accumulating disk matter drops like $t^{-2}$. The fast drop in magnetization leads to a hotter disk and to the reduction of $\tilde\lambda$ below 1.
In the case of a shallower magnetization profile (blue line due to a steeper density profile and green line due to a shallower magnetic field profile), $\tilde\lambda$ remains larger than one for a longer time, leading to a longer lasting jet.

We generally find that when $ \alpha \lesssim 0.5 $ the magnetization profile is steep and the average $ \beta_p $ in the disk rapidly increases. Depending on the magnetic field normalization, we find that the typical jet duration in such cases is $ \lesssim 1 $ s.
Moderate values of $ \alpha \sim 1 $ can support jet launching for $ \teng \gtrsim 1 $ s, but do not seem to reconcile with the typical GRB durations of $ \teng > 10 $ s (see e.g. yellow curve in Fig. \ref{fig:luminosity} where the luminosity drops after $ \sim 2 $ s). Longer duration can be obtained with stronger initial magnetic fields, however in this case the jet luminosity might be too high.
Progenitors with high values of $ \alpha \gtrsim 2 $ seem to be the ones compatible with engine activity of $ \teng \approx \tengmax $, but are ruled out due to excessive luminosity and temporal evolution of the lightcurve (\S\ref{sec:luminosity}).

One possibility of why $ \teng $ is shorter than the observed long GRB durations is differences in the magnetic profile. 
For example a magnetic field that drops radially slower from the core, such as the one used in model $ \alpha0BshLc$, results in a shallower magnetization profile, $ \sigma \propto r^{\alpha-5} $. This can prolong the duration of jet engine activity, thereby bringing progenitors with $ \alpha = 1 $ into agreement with typical GRB durations as well.
Alternatively, if the disk viscosity is high, like in the case of efficient neutrino cooling, $\lambda_{{\rm MRI}}$ may increase and lead to larger scale turbulence and a longer jet launching duration \citep{Masada2007}.
A variety of density and magnetic field profiles among GRB progenitors may play a major part in setting the wide distribution of GRB durations, from bursts that last less than a second to hundreds of seconds.

\subsection{Luminosity}\label{sec:luminosity}

\begin{figure}
		\centering
		\includegraphics[scale=0.205]{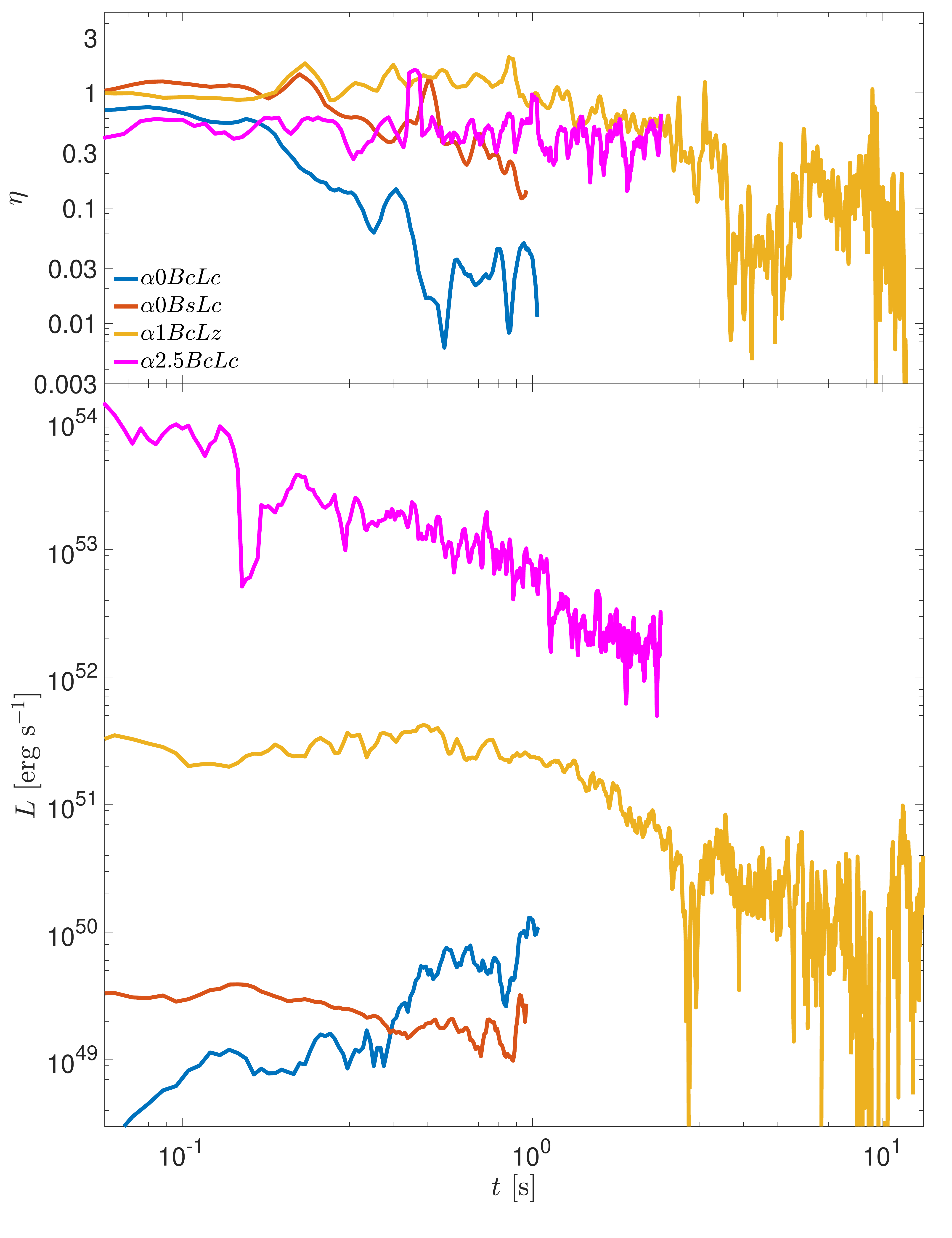}
		\caption[]{Different density profiles lead to different efficiencies (top) and luminosities (bottom), in the models that we specifically designed to be slightly above the critical magnetic flux so they can launch jets. The quantities are measured at the BH horizon as a function of time in a variety of progenitors: three with $ \Phi_h = \Phmin $: $ \alpha = 0 $ (blue), $ \alpha = 1 $ (yellow) and $ \alpha = 2.5 $ (magenta); and one progenitor with $ \Phi_h \approx 3\Phmin $ and $ \alpha = 0 $ (red). We find that over time the fastest growing mode of the MRI, $\lambda_{\rm MRI}$, drops relative to the disk thickness, $H$, resulting in the magnetic flux diffusing outwards and jet efficiency dropping (see \S\ref{sec:engine}).
		The jet power in progenitors with $ \alpha \sim 0 $ or $ \alpha > 2 $ exhibits temporal evolution while $ \eta \sim 1$.		}
		\label{fig:luminosity}
\end{figure}

We find that during the time of a stable jet launching, the jet luminosity satisfies $ L \approx \dot{M}c^2 $, implying that the efficiency remains at $ \eta \gtrsim 0.5 $ (top panel of Fig. \ref{fig:luminosity}), consistent with a MAD state.
However, if the jet operates on the timescale of $ t \lesssim 0.1 $ s, it does not manage to sustain the high-efficiency $\eta\sim1$ state for a long time. This can happen either when $ \beta_p $ is too high so large scale magnetic field structure is absent, or in the case of the weak subrelativistic outflow scenario (see \S\ref{sec:launching}).
The bottom panel of Fig. \ref{fig:luminosity} depicts the jet luminosity for a variety of progenitors.
Since $ L \propto \dot{M} $, the jet luminosity features a global temporal evolution in time, in addition to a short timescale variability due to variations in the accretion power.
The absence of observed temporal evolution in GRB lightcurves\footnote{If the radiation efficiency is of order unity, as suggested by observations, then the GRB light curve behavior reflects the jet power.} (\citealt{McBreen2002a}, but see \citealt{McBreen2002b} for changing lightcurves as $ L \propto t^{\pm 1} $) places a further constraint on the allowed values of $ \alpha $.

Jets in stars with $ \alpha \gtrsim 2 $ (magenta curve in Fig. \ref{fig:luminosity}) are in tension with observations due to both their very high luminosity needed to overcome the high central density ram pressure, and a monotonically decreasing accretion rate that drops over orders of magnitude during the typical duration of long GRBs.
A temporal variation may also arise in progenitors with $ \alpha \approx 0 $, where the accretion mass rate, and subsequently the jet luminosity, increases with time, as can be seen during the first $ \sim 0.5 $ s in model $ \alpha0BcLc $ (blue curve). When the jet is stronger (model $ \alpha0BsLc $, red curve) the suppression of accretion by the cocoon may level the accretion rate such that it is nearly constant.
Stars with $ \alpha \sim 1 $ are found to be the most consistent with observations, maintaining both $ L \approx 10^{51}~{\rm erg~s^{-1}} $ and a rather constant luminosity as long as the engine is highly active (yellow curve) as indicated by observations.
Once the global magnetic field structure around the BH is disrupted, the efficiency drops and the jet gradually shuts off (\S\ref{sec:engine}). This effect occurs mostly in systems with $ \alpha < 1 $ and is manifested as a break in the luminosity and $ \eta $ curves. The break is present at $ t \lesssim 1 $ s when $ \alpha = 0 $, and at $ t \gtrsim 1 $ s when $ \alpha \gtrsim 1 $. The efficiency remains high for a longer time in systems with higher $ \alpha $ as demonstrated in Fig. \ref{fig:luminosity} for the cases of $ \alpha = 2.5 $ (magenta curve) and $ \alpha = 1 $ (yellow curve). Interestingly, observations point at an anti-correlation relationship between the jet luminosity and the GRB peak time, $ L\propto t_{\rm peak}^{-1.52} $ \citep{Dainotti2015}. This could be explained by the temporal evolution of different luminosities, with the luminosity of powerful jets drops with time and that of weak jets grows over time.

\section{Jet evolution}\label{sec:magnetohydrodynamics}

After the jet is launched, it interacts with the infalling dense stellar envelope. The interaction of the jet head with the star shocks the jet and stellar material to form a weakly-magnetized cocoon that collimates the jet. The jet-cocoon-star interplay ultimately regulates the jet evolution in the star. While this is not the main focus of this paper, we report of two major features that are found in our simulations as the first self-consistent 3D GRMHD simulations of the collapsar model. A detailed analysis of the two will be presented in a follow-up work.

\subsection{Tilt of the disk} \label{sec:tilt}
The high pressure that grows in accretion disks leads to release of winds from the equatorial plane towards the polar axis, which facilitate the jet collimation at its base. As the jet propagates farther in the star, its collimation becomes supported by the pressurized back-flowing material of the cocoon.
Our simulations show that the heavy parts of the cocoon, which are close enough to the BH fail to become unbound and fall towards the BH. When such relatively heavy material bumps into the jet, it is deflected sideways and falls onto the accretion disk. If enough angular momentum is carried by such blobs, or if this process reoccurs several times in the same direction, it tilts the disk by virtue of altering its angular momentum, and subsequently tilts the jet launching direction as well.
The relaunching of the jet on an alternative path may considerably prolong its breakout from the star and even result in a failed jet.

\begin{figure}
		\centering
		\includegraphics[scale=0.094]{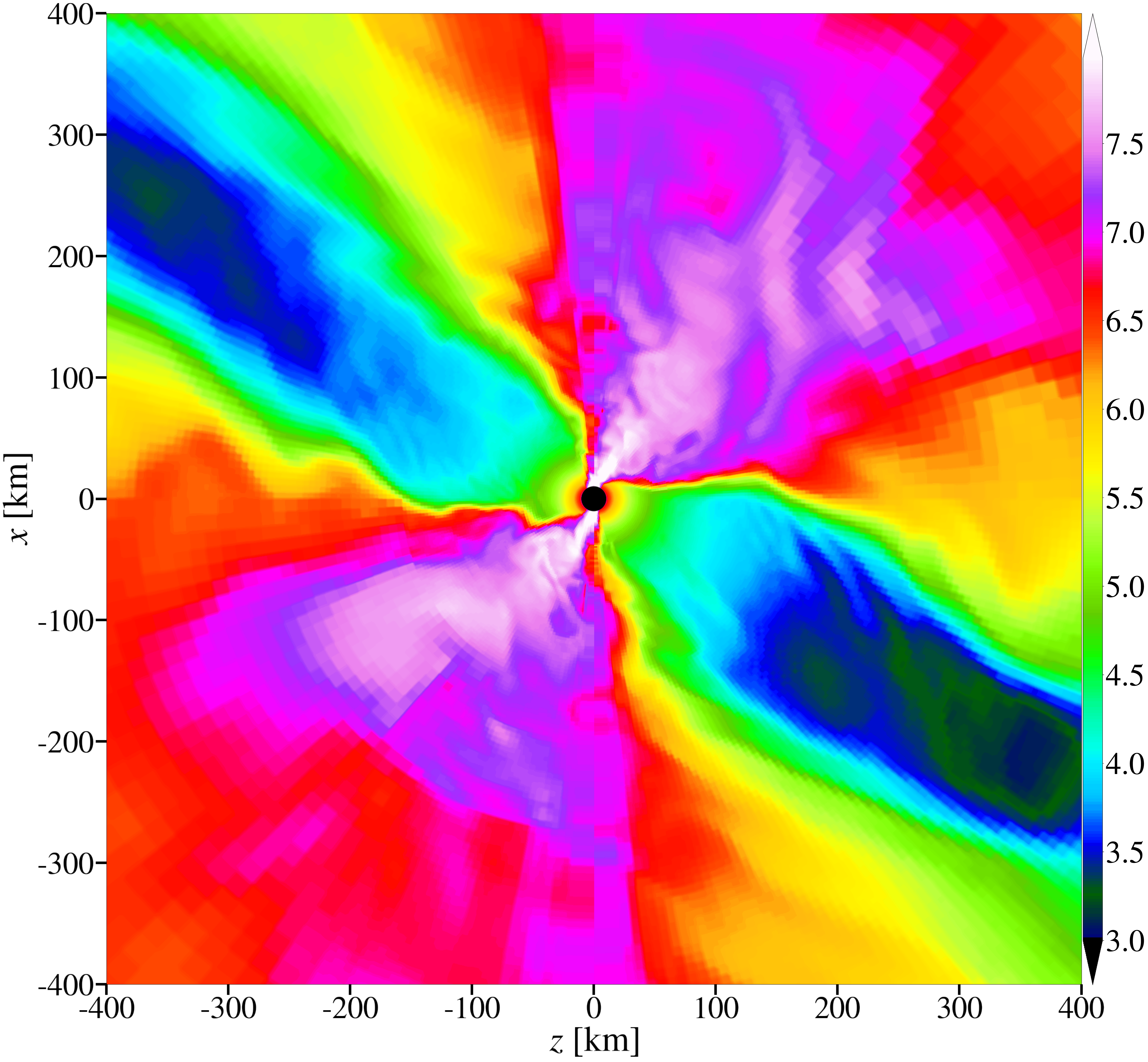}
		\caption[]{Deflection of infalling gas by the jet breaks the symmetry and results in a tilted accretion disk and jet axis.
		Shown is the logarithmic mass density map from model $ \alpha1BcLz $, $ 2.3 $ s after the initial collapse.
		The entire tilt process can be seen in a movie in \url{http://www.oregottlieb/collapsar.html}.
		}
		\label{fig:tilt}
\end{figure}

Fig. \ref{fig:tilt} depicts a zoom-in logarithmic density map of the BH vicinity. It is shown that the disk and the jet are tilted by $ \sim 40^\circ $ (note that a non-tilted disk lies on the $ \hat{x}-\hat{y} $ plane), and in some simulations may reach up to $ \sim 60^\circ $ tilt. If the change in the jet launching direction is substantial, and the time cycle over which it changes is comparable to the GRB duration, the jet head may leave traces of its tilt even after breaking out from the star. Such a process could have profound implications on the expected emission from GRB jets, such as a periodicity in the lightcurve over the precession timescales.

\subsection{Magnetic dissipation} \label{sec:dissipation}

When a Poynting flux dominated jet is collimated, current driven instabilities, most notably kink instability, grow in the jet and dissipate magnetic energy \citep[e.g.,][]{Levinson2013}.
\citet{Bromberg2016} studied the propagation of such jets in stellar envelopes using 3D RMHD simulations. 
The jets were launched by the rotation of a perfectly conducting sphere carrying a monopole magnetic field into a medium with a power-law density profile.
They found that when the jets are collimated, they form narrow nozzles inside which local kink modes grow and dissipate the jet magnetic energy to a level of $ \sigma\simeq1 $. 
Above the nozzle, energy can continue to dissipate via stochastic turbulent motions, though at a lower rate \citep{Bromberg2019}. Such dissipation was not seen in the RMHD simulations of \citet{Bromberg2016}, possibly since the simulation box was too small and the jets were not sufficiently evolved. If magnetic dissipation is sufficient to reduce the magnetization to a level of  $ \sigma \ll 10^{-2} $, the jet continues to evolve as a hydrodynamic jet \citep{Levinson2013,Gottlieb2020b}. In this regime hydrodynamic instabilities growing on the jet boundary will cause strong mixing between jet and cocoon material and may disrupt the jet \citep{Gottlieb2021a}. In the intermediate regime of $ 10^{-2}\lesssim \sigma\lesssim 1 $ jets are stable to both current-driven and boundary instabilities \citep{Gottlieb2020b}. It is therefore important to carefully analyze the jet magnetization above the nozzle and distinguish between the various cases.

\begin{figure*}
		\centering
		\includegraphics[scale=0.12,trim=0 20 0 0]{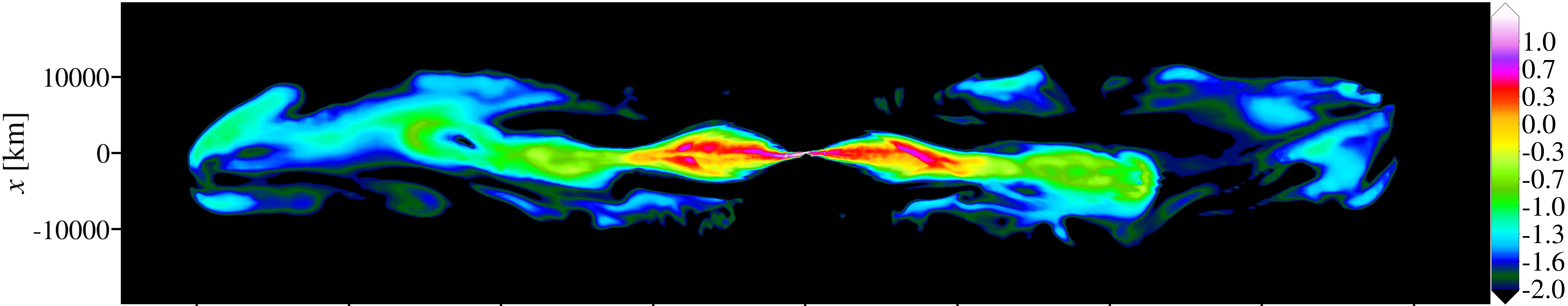}
		\includegraphics[scale=0.2505,trim=56 0 0 0]{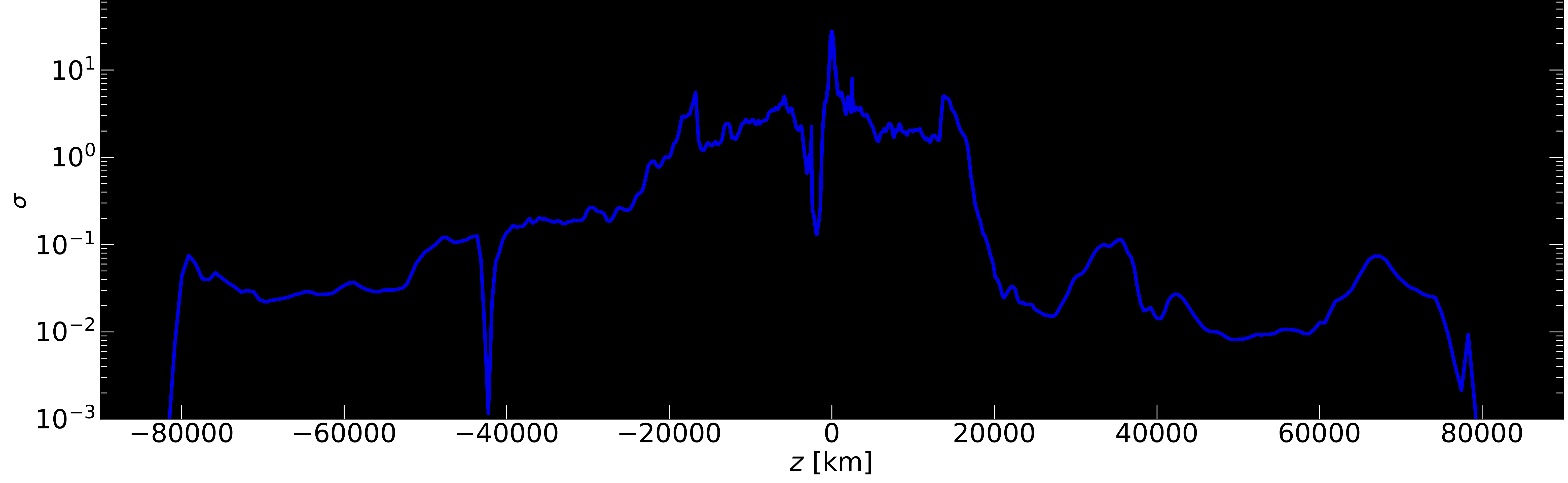}
		\caption[]{Substantial magnetic dissipation at the collimation throat results in a weakly-magnetized jet ($ \sigma \ll 1 $) above the nozzle. Shown is $ {\rm log}(\sigma) $ meridional map of model $ \alpha1BcLc $ and its corresponding magnetization profile along the jet axis at $ t \approx 1.2 $ s.
		}
		\label{fig:sigma}
\end{figure*}

Our GRMHD simulations are different from those of \citet{Bromberg2016} in two important properties. i) we do not assume a magnetic field configuration on the horizon but let it accumulate through accretion. ii) we account for gravity, thus the jet propagates into a medium that is free-falling to the center. We find that qualitatively our results are in agreement with \citet{Bromberg2016}, as the jet plasma exhibits substantial magnetic dissipation after it becomes collimated and passes through the nozzle. The simulations show that the dissipation continues also above the nozzle reducing the magnetic energy to a level of $ 10^{-2} \lesssim \sigma \lesssim 10^{-1} $, as shown in Fig. \ref{fig:sigma} \citep[and was also recently found by][]{Gottlieb2021d}. The reason for the stronger dissipation may lie in the fact that the jets propagate to a much larger distance than previously investigated, thus allowing more time for the stochastic processes to reduce the magnetic field. Another possibility is insufficient resolution, especially at the nozzle region that leads to a high numerical diffusivity and an excess of magnetic energy dissipation. 
During the late stages of the jet evolution the nozzles are resolved by $ \sim 6 $ cells in the lateral direction at their narrowest point. While a similar behavior is detected in a convergence test where the resolution is doubled, a detailed study of this dissipation is needed and will be preformed in a separate work. The question remains whether the properties of the weakly-magnetized jet above the nozzle (e.g., Lorentz factor and energy) are compatible with those inferred from observations.

\section{Implications to breakout \& emission} \label{sec:breakout_emission}

The jets ultimately break out of the star to power the luminous GRB emission.
Based on the jet propagation and properties at the end of the simulations, we derive the conditions needed for the jet to break out within a typical breakout time of $ \tbobs \sim 10 $ s inferred from observations \citep{Bromberg2012}, and discuss the implications for the $ \gamma $-ray signal. In this analysis we ignore potential effects of the tilt on the jet breakout and emission, which are left for a future work.
We stress that the evolution of the jet above the collimation nozzle could be affected by the limit in our simulations of $ \Gamma_\infty \leq 25 $. Under this limit, even moderate mixing with cocoon material that penetrates the jet, may reduce the asymptotic Lorentz factor to $ \Gamma_\infty \lesssim 3 $. Had the jet been launched with $ \Gamma_\infty > 100 $, as inferred from observations, it is possible that it would have maintained a higher asymptotic Lorentz factor after the mixing.

\subsection{Breakout}\label{sec:breakout}

In \S\ref{sec:luminosity} we showed that progenitors with steep density profiles $ \alpha \gtrsim 2 $ are in tension with observations, since the jets that are able to break out of them are too powerful and show temporal evolution that is absent in typical GRB lightcurves. Our simulations also show that the combination of a very strong jet and a steep density profile leads to a fast ($ t \sim 0.1 $ s) acceleration of the jet head to mildly-relativistic velocities. Once the head reaches $ v_h \approx c $, after $ \sim 0.5 $ s, it becomes detached from the engine, namely even if the engine stops at that point, the information will fail to reach the head before it breaks out of the star.
In a population of stars with a distribution of engine activity times, the inferred breakout time from observations will be of the order of this critical time \citep[e.g.][]{Bromberg2012}, implying that $t_b \sim 1~{\rm s} \ll \tbobs $. This supports our previous claim that stars with $ \alpha \gtrsim 2 $ cannot be typical progenitors of long GRBs.
Note that the results above were obtained for $ \Phi_h \approx \Phmin $, and jets with weaker initial magnetic field would fail to produce GRBs.

In density profiles with $ \alpha < 2 $, the jet head typically propagates in subrelativistic velocities inside the stellar envelope. At distances $r\gtrsim10^9$ cm, the jet material close to the head is weakly-magnetized ($ \sigma \ll 1 $), and the free-fall time of the stellar material ahead of the jet is larger than the jet propagation time. We can use the scaling relations obtained for the propagation of hydrodynamic jets in stationary envelopes, $ v_h \propto (t^{\alpha-2}L)^{\frac{1}{5-\alpha}} $, \citep[e.g.][]{Bromberg2011,Harrison2018} to predict the head velocity at large distances from the center. We showed that the jet luminosity roughly follows $ L \propto t^{(2-2\alpha)/3} $, thus
\begin{equation}\label{eq:vh_NR}
v_h \propto t^{\frac{\alpha-4}{3(5-\alpha)}}, 
\end{equation}
i.e. all non-relativistic heads decelerate over time (and not only when $ \alpha < 2$ as in the case of a constant luminosity), with an absolute velocity that depends on $ B_h $ . Using this expression and the jet velocity in our $ \alpha \lesssim 0.5 $ simulations at early times, we deduce the jet velocity at later times\footnote{In these simulations $ t_b \gg 10 $ s, so they do not last until jet breakout.}. In the case of $ \Phi_h \approx 10\Phmin $, integration over the jet velocity yields $ t_b $ of a few dozen seconds. This result is roughly consistent with observations, but requires a different magnetic profile than the core+dipole used in our simulations, that can sustain $ \teng \gg 1 $ s (\S\ref{sec:engine}).
In the progenitor with $ \alpha \approx 1 $ and $ \Phi_h \approx \Phmin $ (models $ \alpha1BcLc, \alpha1BcLz $) the jet propagates at $ v_h \approx 0.15 $ c. Substituting $ \alpha = 1 $ in Eq. \ref{eq:vh_NR} yields $ v_h \propto t^{-1/4}$, which results in an estimated breakout time of $ t_b \approx \tbobs \approx 10 $ s, consistent with the breakout time found in our simulations. In the simulation the weakening of the jet after a couple of seconds implies that $ \teng < 10 $ s (see yellow curve in Fig. \ref{fig:luminosity}), hence in these cases the breakout is of the shocked jet material rather than the jet itself, and may resemble a low-luminosity GRB, similar to the case when the disk is absent. However, a slight modification of $ B_h $ or of the magnetic field profile would enable the jet to work for a few dozen seconds, as suggested by observations.

Our simulations focus on progenitors with $ R_\star = 4\times 10^{10} $ cm, and the propagation and breakout of jets of other, likely larger, $ R_\star $ differs in several ways. It has two constructive effects: it prolongs the breakout time and lower the central mass density (when $ M_\star $ is kept fixed), thereby allowing lower luminosity jets. On the other hand, it also prolongs the accretion process, extending the engine duration as $ \tengmax \propto R_\star^{1.5} $.
However, one should remember that the actual jet duration also depends on the magnetic field structure and the physical properties that take place in the disk, which may differ from the case of a typical WR star. Therefore, the connection between the size of the star and the actual central engine activity time is not straightforward. 
Finally, we note that a cylindrical rotation (model $ \alpha1BcLz $) mitigates the jet propagation at $ r \lesssim 10^9 $ cm through the formation of a low density funnel on the polar axis. At larger radii the timescale over which a significant funnel on the polar axis emerges is longer than the time it takes the jet to reach those radii, owing to the larger angular distance between different angles. Consequently, at $ r \gtrsim 10^9 $ cm the effect of the funnel on the jet propagation, and thus also on the breakout time, is minimal.

\subsection{Prompt emission}

The prompt emission originate from dissipation processes that likely take place in the jet far from the progenitor. The powering mechanism of the emission is a topic of a long-lasting debate and beyond the scope of this paper. However, some general properties which can affect the observed emission can be obtained from our simulations.
Most of the jet magnetic energy dissipates at the collimation nozzle, which is located deep in the star, at a large optical depth ($ \tau \gg 10^3 $), implying that there is enough time to thermalize the photons generated during the dissipation process to produce a rest-frame spectral energy distribution peak at $ \sim 50 $ keV required by observations \citep[e.g.,][]{Ito2020}.
Above the nozzle the energy density is dominated by thermal pressure with $\sigma\approx10^{-2}$, and the jet continues to evolve similar to a hydrodynamic jet. The jet accelerates under its internal energy above the collimation nozzle, which becomes the effective origin from which the jet starts its acceleration\footnote{Note that previous hydrodynamic studies found that in fact the free expansion of the jet begins farther away from the collimation nozzle \citep{Lazzati2009,Gottlieb2019}, such that the photospheric efficiency is even higher.}. This pushes the coasting radius of the jet to distances on the order of the photospheric radius. As a result, an inevitably high photospheric efficiency is expected, implying that even if the jet is launched as Poynting-flux dominated, photospheric emission is a key component in the prompt signal \citep{Gottlieb2019}.

Another common property of GRB lightcurves is the rapid temporal variability \citep[e.g.,][]{Ramirez-Ruiz2000,Nakar2002a,Nakar2002b}. 
The observed fluctuation timescales of $ \sim 0.1 $ s are much shorter than the expected period of precession or global temporal evolution of $ \dot{M} $, which is present if $ \alpha \not\simeq 1 $, thus they are unlikely to be linked to those phenomena.
The rapid variability could originate in mixing between the jet and the stellar material or by the abrupt nature of the central engine. \citet{Gottlieb2021b} found that if the jet is weakly-magnetized above the nozzle and its engine is intermittent, as we find in our simulations, then its observed variability is primarily dictated by the central engine activity. Our simulations confirm this prediction as the central engine intermittency takes place on timescales of a few dozen ms, consistent with the fluctuations in the accretion rate, and with the observed variability of GRB lightcurves \citep{Bhat2013}. The physical mechanism responsible for this timescale is beyond the scope of this paper.
We emphasize that detailed calculations of the photospheric emission are needed for e.g., addressing the light-curve variability and non-thermal spectrum that arise in observations.

\section{Summary \& Discussion}\label{sec:summary}

The collapsar model provides a framework for the observed link between long GRBs and CCSNe of type Ic. According to this model when a rapidly-rotating massive star collapses to a BH, a relativistic jet may be launched, and ultimately power the GRB signal after it breaks out from the star. Although the model provides a reasonable overall picture it does not explain why GRBs are only associated with some CCSNe Ic \citep{Levan2016}. That begs the questions i) what type of progenitors support a relativistic jet launching? ii) What are the necessary conditions that allow for a relativistic jet to break out from the star? In this paper we addressed these questions combining analytic estimates with a set of novel 3D GRMHD simulations of the collapsar model, which can cover the entire jet evolution in the star, from self-consistent launching to breakout. 

Our initial setup consists of a stellar core with a uniform vertical magnetic field surrounded by an envelope carrying a dipolar field.
We generally find that in order to launch a successful jet two requirements must be fulfilled: i) 
the poloidal magnetic field at the horizon needs to be strong enough and have a large coherence length, to power a jet that can overcome the ram pressure of the infalling material on the pole. In our setup this requirement was translated to an initial peak magnetization $ \sigma \gtrsim 10^{-2}$ at the edge of the uniformly magnetized stellar core. In this case, the magnetic field in the accretion disk could be amplified without being disrupted by small scale MRI driven turbulence, allowing a successful launching of the jet. Though the exact value of $\sigma$ may depend on the initial magnetic field profile, the general requirement of a relatively high initial magnetization is likely robust (see below).
ii) the centrifugal barrier of the collapsing matter must be at $ r > \risco $ to allow the formation of an accretion disk.
We found that whether the system satisfies various permutations of the above two requirements may lead to several possible outcomes:
\begin{enumerate}
\item If the angular momentum is large enough, but the magnetic field is too weak, the energy extracted from the BH is insufficient to power a jet. Instead, a growing accretion shock wave forms, energized by the disk wind, the properties of which resemble quasi stationary accretion shocks (SASI) in SNe.
The emergence of SASI under these conditions suggests that the origin of SN/SN+GRB dichotomy in CCSN Ic stars may lie in the magnetic energy of the star.
\item If the magnetic field is strong but the angular momentum is too low, an accretion disk fails to form and field amplification does not occur. In this case, a brief weak jet is launched due to compressed field that accumulates on the horizon. Once the field reconnects, the jet shuts off. Energy continues to be extracted via weak fields resulting in a subrelativistic outflow whose dynamics depends on the density profile in the star. It expands quasi-spherically in a flat density profile, whereas in steep density profiles the outflow can accelerate and eventually break out with properties similar to that of a low-luminosity GRB, implying llGRBs may originate from low angular momentum stars.
\item If both of the above requirements are fulfilled, a relativistic BZ jet is launched. We found that during the lifetime of the jet its efficiency remains close to unity, implying that the launching occurs when the system is at a MAD state. In this case, the jet power is essentially the accretion power, which can be approximated as the mass flow from a spherical free-fall collapse, up to a small correction factor.
If the magnetization profile is too steep, the central engine activity is shorter than the time it takes for the jet to reach the surface of the star and break out of it. In this case, the breakout is of the shocked jet material, which may also be associated with llGRBs.
\end{enumerate}

The collapse of the stellar material along the future jet axis enforces a minimal jet luminosity, $\Lmin$ to ensure the survival of the jet. The value of $\Lmin$ depends on the density profile and can be used to constrain the parameter space to values allowed by observations. The jet launching duration is found to be dependent on the initial magnetization profile in the star, with lower magnetization reducing the MRI wavelength and leading to an incoherent magnetic field structure that does not support continuous jet launching.
Within our set of models with power-law inner post-BH formation density profiles, $\rho\propto r^{-\alpha}$, we find that stars with $ 0.5 \lesssim \alpha \lesssim 1.5 $ may be responsible for the entire range of long GRB observables. The observed jet power, which spans two orders of magnitude from  $ \sim 10^{49}~\rm {erg~s^{-1}}$ to $ \sim 10^{52}~{\rm erg~s^{-1}} $, can be partly attributed to different $ \alpha $ in this range. The relation between the accretion rate and the luminosity also implies that the lightcurve shows only a weak evolution in time. While in our $ \alpha = 1 $ progenitors the jet duration was too short to allow a relativistic jet breakout, the observed GRB duration which ranges between sub-second to tens of seconds, can be attained for different combinations of magnetic field and density profiles.
In the case of progenitor with $ \alpha \gtrsim 2 $, the minimum required jet power is $ \Lmin > 10^{52}~{\rm erg~s^{-1}} $, the lightcurve shows evolution in time, and the breakout time is $ t_b \sim 1 $ s, all of which are inconsistent with observations. Another constraint on the properties of the progenitor star may come from the requirement of minimal peak magnetization of $\sigma\gtrsim10^{-2}$ at the edge of the magnetized core. For our set of models that support a successful jet launching ($0.5\lesssim\alpha\lesssim1.5$), this translates into an initial core magnetic field of $\sim5\times10^{11}$~G to $\sim5\times10^{12}$~G, or to a surface dipolar field of $\sim10-100$ kG. We summarize our findings in Tab. \ref{tab:model_constraints}).

\begin{table}
	\setlength{\tabcolsep}{10.7pt}
	\centering
	\renewcommand{\arraystretch}{2}
	\begin{tabular}{c|c | c | c}
		
		& $ \alpha \lesssim 0.5 $ & $ 0.5 \lesssim \alpha \lesssim 1.5 $ & $ \alpha \gtrsim 2 $
		\\	\hline
        ${\rm log}(\Lmin~[{\rm erg~s^{-1}}$]) & $ \lesssim 49 $ & $ 49-52 $ & \color{red} $ \gtrsim 52 $
        \\
        $ L \propto t^{\zeta} $ & \color{red} $ \zeta \gtrsim 0 $ & $ \zeta \sim 0 $ & \color{red} $ \zeta \lesssim 0 $
        \\ 
        $ B(r>r_c)\propto r^{-\xi} $ & $ \xi \lesssim 1.5 $ & $ \xi \lesssim 2 $ & $ \xi \lesssim 2.5 $
	\end{tabular}

	\caption{Inferred physical quantities for different post-BH formation inner density profile of the progenitor. Cells marked in red are inconsistent with observations.  The values of $ \xi $ are constrained by the requirement for sufficiently long engine working times.
		}
\label{tab:model_constraints}
\end{table}

Our simulations show significant magnetic energy dissipation above the collimation nozzle to a level of $\sigma\approx10^{-2}$, which has important implications to post-breakout jets:
i) most of the photons required to keep the jet internal energy at $\sim50$ keV can be produced during the dissipation processes that take place at the nozzle and somewhat above it \citep{Gottlieb2019}. ii) the nozzle is the effective origin of the nearly hydrodynamic jet that emerges from it, and sets the location of the coasting radius close to the photosphere to allow an efficient photospheric emission \citep{Gottlieb2019}. iii) The observed variability in the prompt emission is most likely related to temporal variations in the launching process and not to hydrodynamic instabilities at the jet boundary, as the latter are inhibited in the presence of weak magnetic fields \citep{Gottlieb2021b}. We stress however that further study is required to validate the low magnetization of jets at large distances.

Interestingly, we find that fallback of cocoon (shocked stellar and jet contents) material onto the disk may apply torques to the disk. These torques tilt the disk-jet system and thereby alter the direction of jet launching. This may lead to some interesting observational consequences. 
First, the tilt results in relaunching of the jet in a different direction. This prolongs the breakout time of the jet from the star and may even cause it to fail entirely.
Second, if the jet emerges from the star at different directions, this would utterly change our understanding of the prompt emission mechanism and the statistics of long GRBs.
Third, if the typical time over which the tilt develops is longer than the breakout time, the jet may experience significant precession. The period of the procession could be reflected by periodicity in the lightcurve. That could explain the intermediate GRB timescale of $ \sim 1 $ s, as a third characteristic timescale, in addition to the variability timescale and the total burst duration \citep{Nakar2002a}.

Our simulations do not include neutrino transport and this may have several important effects on our findings. First, neutrino emission cools the disk and may alter the jet properties accordingly, e.g., change the length scale of the magnetic field accreted onto the BH and affect the jet duration. Second, SNe simulations show that neutrino losses play a major role in stalling the accretion shock wave, and thus have to be included to properly model the SASI evolution. Third, while neutrino-antineutrino annihilation \citep{Eichler1989,Popham1999,MacFadyen1999}
might fall short in powering GRB jets of typical power by virtue of producing pressure-driven pairs, they may form a low density funnel that mitigates the BZ jet to breach through the stellar envelope. However a polar funnel, which is also present in progenitors with rotation velocity constant on cylinders (see \S\ref{sec:launching}), is not anticipated to alter our conclusions considerably. We will target the neutrino effects on the collapsar model in a future work.

\section*{Acknowledgements}
	
We thank Eliot Quataert and the anonymous referee for useful comments.
OG is supported by a CIERA Postdoctoral Fellowship. OB acknowledges support by an ISF grant 1657/18 and by an ISF (Icore) grant 1829/12. OB and AT were also partly supported by an NSF-BSF grant 2020747. AT was supported by NSF grants
AST-2107839, 
AST-1815304, 
AST-1911080, 
AST-2031997. 
An award of computer time was provided by the Innovative and Novel Computational Impact on Theory and Experiment (INCITE) program under award PHY129. This research used resources of the Oak Ridge Leadership Computing Facility, which is a DOE Office of Science User Facility supported under Contract DE-AC05- 00OR22725.
The authors acknowledge the Texas Advanced Computing Center (TACC) at The University of Texas at Austin for providing HPC and visualization resources that have contributed to the research results reported within this paper via the LRAC allocation AST20011 (\url{http://www.tacc.utexas.edu}).
This research was also enabled in part by support provided by Compute Canada allocation xsp-772 (\url{http://www.computecanada.ca}).

\section*{Data Availability}
	
The data underlying this article will be shared on reasonable request to the corresponding author.

\bibliographystyle{mnras}
\bibliography{ref}

\label{lastpage}
\end{document}